\documentclass{article}

\usepackage[utf8]{inputenc} 
\usepackage[T1]{fontenc}    
\usepackage{hyperref}       
\usepackage{url}            
\usepackage{booktabs}       
\usepackage{amsfonts}       
\usepackage{nicefrac}       
\usepackage{microtype}      
\usepackage{lipsum}
\usepackage{fancyhdr}       
\usepackage{graphicx}       
\graphicspath{{media/}}     
\usepackage{xcolor}
\usepackage{xspace}
\usepackage{subcaption}
\usepackage{multirow}
\usepackage{rotating,tabularx}
\usepackage[numbers]{natbib}
\usepackage{amsmath}
\usepackage{cleveref}
\usepackage[scientific-notation=true]{siunitx}
\usepackage{stringstrings}

\usepackage[acronym]{glossaries}

\newacronym{aapm}{TASC}{AAPM Thoracic Auto-Segmentation Challenge}
\newacronym{cranialvault}{Cranial Vault}{Multi-atlas Labeling Beyond the Cranial Vault challenge}
\newacronym{csi}{CSI}{Computational Methods and Clinical Applications for Spine Imaging}
\newacronym{eligibson}{MARSS}{Multi-organ Abdominal CT Reference Standard Segmentations}
\newacronym{lits}{LiTS}{Liver and Liver Tumor Segmentation challenge}
\newacronym{kits19}{KiTS19}{2019 Kidney Tumor Segmentation challenge}
\newacronym{structseg}{StructSeg2019}{Automatic Structure Segmentation for Radiotherapy Planning challenge}
\newacronym{sliver07}{Sliver07}{Segmentation of the Liver 2007 challenge}
\newacronym{verse19}{VerSe19}{Large Scale Vertebrae Segmentation challenge}

\newcommand\expe[1]{Exp\substring{#1}{1}{2}\substring{#1}{3}{$}}

\newcolumntype{C}[1]{>{\centering\arraybackslash}p{#1}}

\newcommand{\organrot}[1]{\rotatebox{90}{#1}}

\newcommand{\structsegurl}{\url{https://structseg2019.grand-challenge.org/}}

\title{Transfer learning from a sparsely annotated dataset of 3D medical images}

\author{
Gabriel Efrain Humpire-Mamani$^{1}$, Colin Jacobs$^{1}$, Mathias Prokop$^{1}$,\\
Bram van Ginneken$^{1,2}$ and Nikolas Lessmann$^{1}$\vspace{0.2cm}\\
  \small $^1$\textit{Diagnostic Image Analysis Group, Radboud University Medical Center,}\\
  \small \textit{Nijmegen, The Netherlands}\vspace{0.05cm}\\
  \small$^2$\textit{Fraunhofer MEVIS, Bremen, Germany}\vspace{0.2cm}\\
  \small \texttt{g.humpiremamani@radboudumc.nl} \\
}
\date{\vspace{-5ex}}

\begin{document}
\maketitle

\begin{abstract}
Transfer learning is a technique used in machine learning where features learned by a model on a large annotated dataset are transferred and leveraged when training a new model for other tasks. This technique can save substantial time and computational resources compared to training models from scratch. In addition, performance may also improve when leveraging pre-trained features. 
Due to the lack of large datasets in the medical imaging domain, transfer learning from one medical imaging model to other medical imaging models has not been widely explored. 
This study explores the use of transfer learning to improve the performance of deep convolutional neural networks for organ segmentation in medical imaging.
A base segmentation model (3D U-Net) was trained on a large and sparsely annotated dataset; its weights were used for transfer learning on four new down-stream segmentation tasks for which a fully annotated dataset was available. We analyzed the training set size's influence to simulate scarce data.
The results showed that transfer learning from the base model was beneficial when small datasets were available, providing significant performance improvements; where fine-tuning the base model is more beneficial than updating all the network weights with vanilla transfer learning.
Transfer learning with fine-tuning increased the performance by up to 0.129 (+28\%) Dice score than experiments trained from scratch, and on average 23 experiments increased the performance by 0.029 Dice score in the new segmentation tasks.
The study also showed that cross-modality transfer learning using CT scans was beneficial.
The findings of this study demonstrate the potential of transfer learning to improve the efficiency of annotation and increase the accessibility of accurate organ segmentation in medical imaging, ultimately leading to improved patient care.
We made the network definition and weights publicly available to benefit other users and researchers.
\end{abstract}

\section{Introduction}

Transfer learning is a widely used strategy when developing image analysis models based on deep learning. The weights of a model trained on a large dataset are transferred to a new model that learns a different task. This transfer initializes the weights of the new model with well-converged and meaningful filters, which gives the new model a head start compared with random initialization and especially improves the performance of models trained with small datasets \citep{Shin16}. While transfer learning is therefore useful in many computer vision and medical image analysis applications, the vast majority of pretrained models available for transfer learning are models trained with two-dimensional non-medical images, such as the ImageNet dataset \citep{Deng09}. Transfer learning from these models can lead to adverse results when training medical image analysis models \citep{Ragh19a}. However, models pretrained on a medical dataset with a large number of three-dimensional medical images and a diverse set of labeled objects are currently not readily available.

Although various annotated datasets with three-dimensional medical images are available under licenses that permit their use for pretraining, most of these datasets are small, and annotations are usually sparse. In virtually all segmentation datasets, only a single or a few anatomical structures are delineated because manual segmentation in three-dimensional datasets is time-consuming and expensive. Most projects are also focused on a specific task that does not require exhaustive annotation of all structures. However, datasets in which many visible anatomical structures are delineated would be best suited for training generic models for transfer learning.

There have been a few efforts to assemble datasets with a larger number of structures annotated, such as the VISCERAL dataset \citep{Jime16}. Other efforts focused on combining datasets and expanding annotations to additional structures, such as \citet{Gibs18b}, who combined two publicly available datasets \citep{Roth15d,Land17} and expanded the annotations to 14 structures. However, increasing the number of annotated structures in a set of scans usually comes at the expense of the number of scans in the dataset.

This paper explores an alternative strategy for training a generic base model for segmentation tasks in medical images that does not require a fully annotated dataset. Instead, we propose to train the base model with a large but sparsely annotated dataset. This dataset is assembled from multiple publicly available datasets with CT images and reference segmentations of various anatomical structures. While each image has at least one delineated structure, we relax the requirement that all structures be delineated in all images and propose a method for training a deep neural network with this kind of sparse annotation. We investigate whether using this base model to initialize the network for a new segmentation task improves the performance. We evaluate whether the size of the training dataset for the new tasks is related to the efficacy of transfer learning. To enable others to use this base model for transfer learning, the network is based on the commonly used 3D U-Net architecture \citep{Cice16}, and the code and weights are made available online.\footnote{\url{https://github.com/DIAGNijmegen/MedicalTransferLearning3D-UNet}}


\section{Related work}

\subsection{Transfer learning}
While transfer learning is widely applied to 2D data, it is not commonly applied in 3D medical imaging due to the lack of large 3D datasets.
Multiple large 2D fully-annotated datasets (ImageNet, MS-COCO, and CIFAR) and spatiotemporal 2D datasets, such as Kinetics, are available in comparison to the small 3D medical imaging datasets~\citep{Shie15}.

Regardless of the image domain, methods trained on 2D images used transfer learning to 3D images ~\citep{Rajp20,Conz20,Yang21,Opbr15,Carn15,Ravi16a} by decomposing the 3D image into a sequence of 2D images.
In \citet{Conz20}, the features of a pretrained network on ImageNet segmented healthy and unhealthy shoulder muscles in MRI using transfer learning.
Similar to action recognition in videos, a pretrained network on Kinetics (large dataset for action recognition in videos) initialized a 3D network to diagnose appendicitis in CT scans~\citep{Rajp20}. The network expects a sequence of frames to recognize an action in a video; the sequence of frames was replaced by a sequence of 2D slices to recognize abnormal regions in a CT scan.
In \citet{Yang21}, 2D pretrained networks are converted to 3D networks; this approach benefits from the large-scale 2D datasets and the 3D context that 3D networks offer.

Although different-domain transfer learning methods showed higher results than methods trained from scratch, same-domain transfer learning showed more reliable results for medical imaging~\citep{Chep19a,Roth18,Zhou21,Chen19a,Ji22}.
\citet{Roth18} trained a cascade of 3D U-Nets using a fully annotated dataset to make 3D medical imaging more accessible for other researchers.
In \citet{Zhou21}, a self-supervised learning method learns from unannotated medical data (LUNA16 dataset without annotations) to obtain a pretrained model, which can be fine-tuned for classification and segmentation.
The method uses an encoder-decoder architecture to perform medical image restoration, which learns the texture and features of organs.
Similar to our approach, \citet{Chen19a} joined data from three medical challenges (MRI and CT) to compose a partially annotated dataset and trained 3D convolutional networks for segmentation and classification.
In a recent study, \citet{Ji22} trained a nnU-Net on a large-scale medical dataset for organ segmentation; transfer learning from that model increased the performance in unseen segmentation tasks from the Medical Segmentation Decathlon  challenge\cite{Anto22}.
The network weights obtained by \citet{Roth18,Chen19a}, and \citet{Zhou21} were released and are publicly available.
Additionally, our method uses a partially annotated dataset composed of 6 publicly available datasets.

\subsection{Learning from sparsely annotated data}
Combining multiple medical annotated datasets could create a large but partially annotated dataset; this data cannot be directly used by methods that depend on fully annotated datasets.
Only few researchers focused on medical image segmentation with partially annotated datasets where methods obtain pseudo-labels from unlabeled images to train networks.
Pseudo-labels can be obtained by approximating the shape and position of a missing label \citep{Dong20,Guo22,Lian23}, relabeling \citep{Peti18}, weak annotations \citep{Tajb20,Wang23}, or by adding constraints such us anatomical prior organ size \citep{Zhou19}. 
By adapting the cross-entropy to learn more from the foreground than the background, \citet{Jin17} trained a 3D network to add more airway branches to the airway segmentation obtained by a previous method.
Multi-stage approaches (i.e., per groups of fully annotated data) served as a multi-organ segmentation network \citep{Fang20,Peti18,Chen19a}.
In an end-to-end solution, \citet{Shi21} proposes two losses to train a network using a partially annotated dataset (data from multiple datasets).
The first loss merges all unlabeled as a single label and the second loss assumes organs are non-overlapped to differentiate between labeled organs and estimated predictions.
\citet{Liu21} used incremental learning to train the network on a different organ in each stage. After four stages, the network segments the organs from four datasets. The method uses a corrective loss to remove low-confidence output.
\citet{Zhan21} proposed DoDNet, a network that emulates a multi-head network (each head for a different task) by proposing a dynamic single-head network.

\section{Dataset}

We combined a number of datasets to assemble a large but sparsely annotated training set for the generic base model, and multiple additional datasets for transfer learning experiments. See \Cref{fig:datasetsforsparselyannotateddataset} for examples of some of the datasets. Many of these datasets have previously been made publicly available as part of segmentation challenges. Because the field of medical imaging is so broad with various modalities such as CT and MRI and various imaging modes per modality such as contrast-enhanced or ECG-gated CT imaging, we limited our experiments to the datasets consisting of thoracic and abdominal CT scans. The evaluation also includes a transfer learning experiment with an MR dataset to investigate whether cross-modality transfer learning is effective. We used the following datasets in our study:

\begin{figure*}[ht!]
  \begin{subfigure}[b]{0.30\textwidth}
    \includegraphics[width=\textwidth]{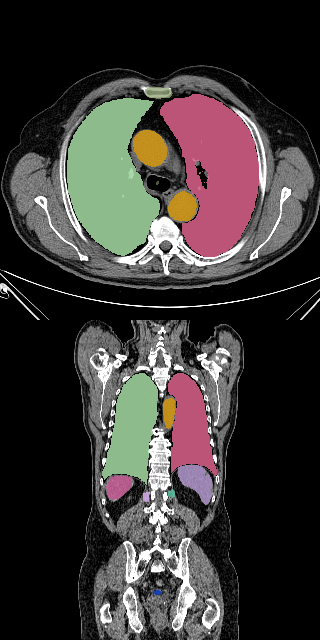}
    \caption{Visceral}
  \end{subfigure}\hfill%
  \begin{subfigure}[b]{0.30\textwidth}
    \includegraphics[width=\textwidth]{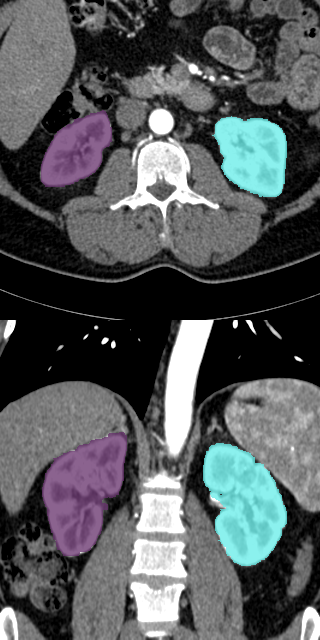}
    \caption{\acrshort{kits19}}
  \end{subfigure}\hfill%
  \begin{subfigure}[b]{0.30\textwidth}
    \includegraphics[width=\textwidth]{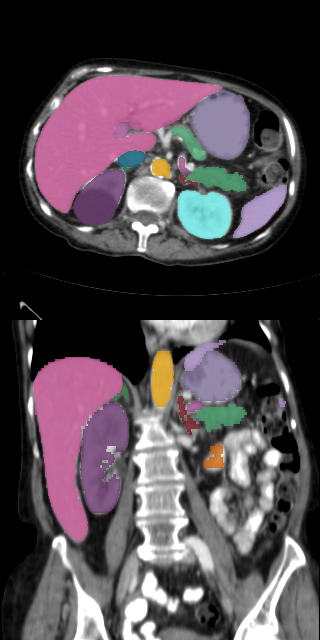}
    \caption{\acrshort{eligibson}}
  \end{subfigure}\\
  \begin{subfigure}[b]{0.30\textwidth}
    \includegraphics[width=\textwidth]{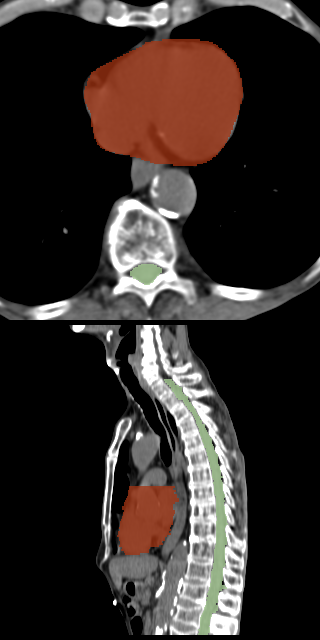}
    \caption{\acrshort{aapm}}
  \end{subfigure}\hfill%
  \begin{subfigure}[b]{0.30\textwidth}
    \includegraphics[width=\textwidth]{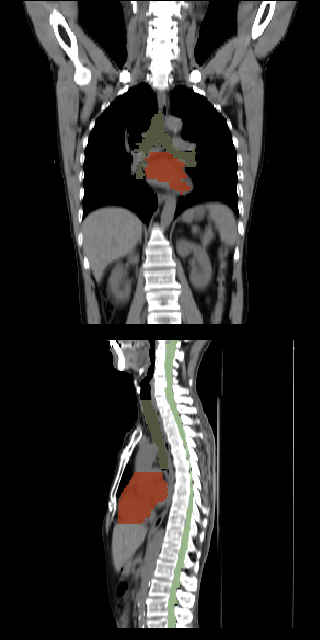}
    \caption{\acrshort{structseg}}
  \end{subfigure}\hfill%
  \begin{subfigure}[b]{0.30\textwidth}
    \includegraphics[width=\textwidth]{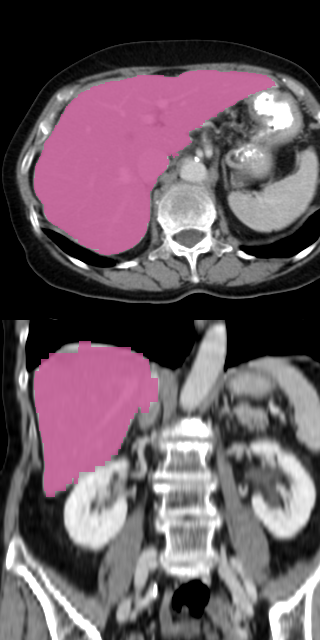}
    \caption{\acrshort{lits}}
  \end{subfigure}
  \caption{Examples CT scans of publicly available datasets. Colored regions represent annotated (reference masks) regions per dataset. The joined annotations of all those datasets compose the partially annotated dataset used for the training of the base model (Exp0). This figure uses the lung window level.}
  \label{fig:datasetsforsparselyannotateddataset}
\end{figure*}

\begin{itemize}
\item The \Gls{kits19} dataset \citep{Hell21} comprises 300 abdominal CT scans from a single medical center. In all scans, the kidneys and kidney tumors were manually delineated and post-processed to remove fat tissue. We included only the 210 scans that were originally made available for model training and validation.
\item The \Gls{lits} dataset \citep{Bili23} comprises 200 thoracic-abdominal CT scans from several medical centers. In the scans, the liver and liver tumors were manually delineated. We included only the 131 scans that were originally made available for model training and validation.
\item The \Gls{eligibson} dataset \citep{Gibs18b} is itself based on two other datasets, namely 47 images from the \acrlong{cranialvault} \citep{Land17} and 43 images from The Cancer Image Archive Pancreas-CT dataset \citep{Roth15d}. In these 90 abdominal CT scans, a total of 14 structures were manually delineated, but 6 of them were only in the \acrshort{cranialvault} dataset (see \Cref{tab:datasetsorgans}).
\item The dataset from the \gls{structseg}\footnote{\structsegurl} contains CT scans of the head and neck and the thorax. We used only the 50 thoracic CT scans from the ``organ-at-risk'' segmentation subtask, for which delineations of the lungs, heart, esophagus, and spinal cord are available.
\item The \Gls{aapm} dataset \citep{Yang18a} comprises 36 thoracic CT scans with delineations of the esophagus, heart, lungs, and spinal cord.
\item The Visceral dataset \citep{Jime16} comprises 40 thorax-abdomen CT scans with delineations of 20 structures (see \Cref{tab:datasetsorgans}). We disregarded four of these structures: the left and right rectus abdominis muscles because they are thin and often difficult to segment structures, the thyroid because we found that the segmentations were of lower quality compared with other structures, and the L1 vertebrae since only a single vertebra was delineated while we used other datasets (see below) with segmentations of all visible vertebrae.
\item The dataset from the 2014 vertebra segmentation challenge at the \gls{csi} workshop comprises 20 spine-focused thorax-abdomen CT scans with delineations of the thoracic and lumbar vertebrae \citep{Yao16}.
\item The \Gls{verse19} dataset \citep{Seku21,Loff20} comprises 180 CT scans of the spine, including thoracic and abdominal scans but also cervical spine scans. We used the 80 scans from the first two training batches and the corresponding delineations of all visible vertebrae.
\item The COPDGene study \citep{Rega10} is a clinical trial that enrolled 10,000 patients with mild to severe COPD who received a thoracic CT scan in one of 21 medical centers in the United States. We used a randomly selected subset of 100 CT scans for which we had access to delineations of the pulmonary lobes.
\item The PROMISE12 challenge dataset \citep{Litj14} comprises 50 T2-weighted MR scans with delineations of the prostate.
\end{itemize}

\noindent
\Cref{tab:datasetsorgans} provides an overview of the number of scans and the annotated anatomical structures in the individual datasets.

\begin{table*}[ht!]
	\caption{Summary of publicly available datasets considered in this paper and the number of CT scans annotated per dataset and organ. The datasets above the middle line (except annotations of the esophagus) served to compose the partially annotated dataset to train Exp0. The datasets below the middle line (+ subset of masks of the esophagus) are fully annotated datasets and were used to analyze the influence of transfer learning on new segmentation tasks.}
    \label{tab:datasetsorgans}
    \renewcommand{\arraystretch}{1.25}
    \setlength{\tabcolsep}{0pt}
    \centering
    \scriptsize
    \begin{tabular}{p{2.3cm}| C{0.45cm} C{0.45cm}	C{0.45cm}	C{0.45cm}	C{0.45cm}	C{0.45cm}	C{0.50cm}	C{0.45cm}	C{0.55cm}	C{0.45cm}	C{0.55cm}	C{0.55cm}	C{0.55cm}	C{0.45cm}	C{0.45cm}	C{0.45cm}	C{0.45cm}	C{0.45cm}	C{0.45cm}	C{0.45cm}	C{0.45cm}	C{0.45cm}	C{0.45cm}	C{0.45cm}	C{0.45cm}	C{0.45cm}	C{0.45cm}	C{0.45cm}	C{0.45cm}	C{0.45cm}}
        \hline
\multirow{2}{1cm}{\textbf{Dataset}} & \multicolumn{30}{c}{\textbf{Organ/structure name}}\\ \cline{2-31}
&	\organrot{Left adrenal gland} &	\organrot{Right adrenal gland} &	\organrot{Aorta} &	\organrot{Bladder} &	\organrot{Duodenum} &	\organrot{\textbf{Esophagus}} &	\organrot{Gallbladder} &	\organrot{Heart} &	\organrot{Left kidney} &	\organrot{Right kidney} &	\organrot{Liver} &	\organrot{\textbf{Lung lobes}} &	\organrot{Left lung} &	\organrot{Right lung} &	\organrot{Pancreas} &	\organrot{Portal \& splenic vein} &	\organrot{\textbf{Prostate}} &	\organrot{Left psoas major} &	\organrot{Right psoas major} &	\organrot{Left rectus abdominis} &	\organrot{Right rectus abdominis} &	\organrot{Spinal cord} &	\organrot{Spleen} &	\organrot{Sternum} &	\organrot{Stomach} & \organrot{Thyroid} & 	\organrot{Trachea} &	\organrot{\textbf{Vertebrae}} &	\organrot{Vertebra L1} & \organrot{Vena cava} \\ \hline
\Gls{kits19} &	 &	 &	 &	 &	 &	 &	 &	 &	210 &	210 &	 &	 &	 &	 &	 &	 &	 &	 &	 &	 &	 &	 &	 &	 &	 &	 &	 & & & \\
\Gls{lits} &	 &	 &	 &	 &	 &	 &	 &	 &	 &	 &	131 &	 &	 &	 &	 &	 &	 &	 &	 &	 &	 &	 &	 &	 &	 &	 &	 & & & \\
\Gls{eligibson} &	47 &	47 &	47 &	 &	90 &	\textbf{90} &	90 &	 &	90 &	47 &	90 &	 &	 &	 &	90 &	47 &	 &	 &	 &	 &	 &	&   90 &	 &	90 &	 &	 &	& & 47	\\
\Gls{structseg} &	 &	 &	 &	 &	 &	\textbf{50} &	 &	50 &	 &	 &	 &	 &	50 &	50 &	 &	 &	 &	 &	 &	 &  &	50 &	 &	 &	 &	& 50 &	 &	& \\
\Gls{aapm} &	 &	 &	 &	 &	 &	\textbf{36} &	 &	36 &	 &	 &	 &	 &	36 &	36 &	 &	 &	 &	 &	 &	 &	&   36 &	 &	 &	 &	 &	 &	& & \\
Visceral &	30 &	27 &	40 &	39 &	 &	 &	38 &	 &	40 &	40 &	40 &	 &	40 &	40 &	38 &	 &	& 40 &	40 &	40 &	40 &	 &	40 &	40 &	 &	32 & 40 &	 & 40 & \\ \hline
\Gls{csi} &	 &	 &	 &	 &	 &	 &	 &	 &	 &	 &	 &	 &	 &	 &	 &	 &	 &	 &	 &	 &	 &	 &	 &	 &	 &	&  & \textbf{20} & & \\
\Gls{verse19} &	 &	 &	 &	 &	 &	 &	 &	 &	 &	 &	 &	 &	 &	 &	 &	 &	 &	 &	 &	 &	 &	 &	 &	 &	 &  & & \textbf{80} & & \\
COPDgene &	 &	 &	 &	 &	 &	 &	 &	 &	 &	 &	 &	\textbf{100} &	 &	 &	 &	 &	 &	 &	 &	 &	 &	 &	 &	 &	 &	 &	 &	& &\\
PROMISE12 (MR)  &	 &	 &	 &	 &	 &	 &	 &	 &	 &	 &	 &	 &	 &	 &	 &	 &	 \textbf{50} &	 &	 &	 &	 &	 &	 &	 &	 &  &	& & &\\
\hline
\textit{Base model} &	77 &	74 &	87 &	39 &	90 & --	 &	128 &	86 &	340 &	297 &	261 &	--  &	126 &	126 &	128 &	47 & -- &	40 &	40 &	-- &	-- &	86 &	130 &	40 &	90 & -- & 90 &	-- & -- &	47 \\ \hline
	\end{tabular}
\end{table*}

\subsection{Harmonization of the reference delineations}
In the large dataset that we created by combining multiple datasets, each dataset contains several structures annotated in more than one of the source datasets. The annotation protocols sometimes differed slightly, and we, therefore, post-processed some of the annotations to harmonize the reference data.

Three datasets delineated the kidneys: \gls{eligibson}, visceral, and \gls{kits19}. In the \gls{kits19} dataset, kidney tumors were annotated with a separate label, while this was not the case in the other two datasets. We, therefore, decided not to consider kidney tumors a separate class and added them to the kidney class in the \gls{kits19} dataset. Additionally, the \gls{kits19} dataset does not distinguish between left and right kidneys, while the other two kidney datasets use two different labels. Therefore, we identified the left and right kidneys in the \gls{kits19} dataset and labeled them the same way as in the other two datasets.

Similarly, the \gls{lits} dataset contains delineations of the liver as well as liver tumors, while other datasets did not separately label liver tumors (\gls{eligibson} and visceral). Therefore, we combined the liver and liver tumor labels in the \gls{lits} dataset.

In the \gls{csi} and \gls{verse19} datasets, we simplified the vertebra segmentation task by removing the anatomical identification task. Instead, we assigned all vertebrae to the same class. This also helped to eliminate any labeling discrepancies between the datasets.

\subsection{Base model and transfer learning tasks}
\label{sec:datadistribution}

The combined dataset contains a total of 736 CT scans and 50 MR scans with delineations of 26 anatomical structures, where each structure is annotated in at least 39 and in up to 340 scans (see \Cref{tab:datasetsorgans}). For the generic base model training set, we selected 22 of these structures; this corresponds to a set of 556 CT scans of which 90\% were used for training and 10\% for validation.

\begin{figure*}[t]
    \centering
    \includegraphics[width=1.0\textwidth]{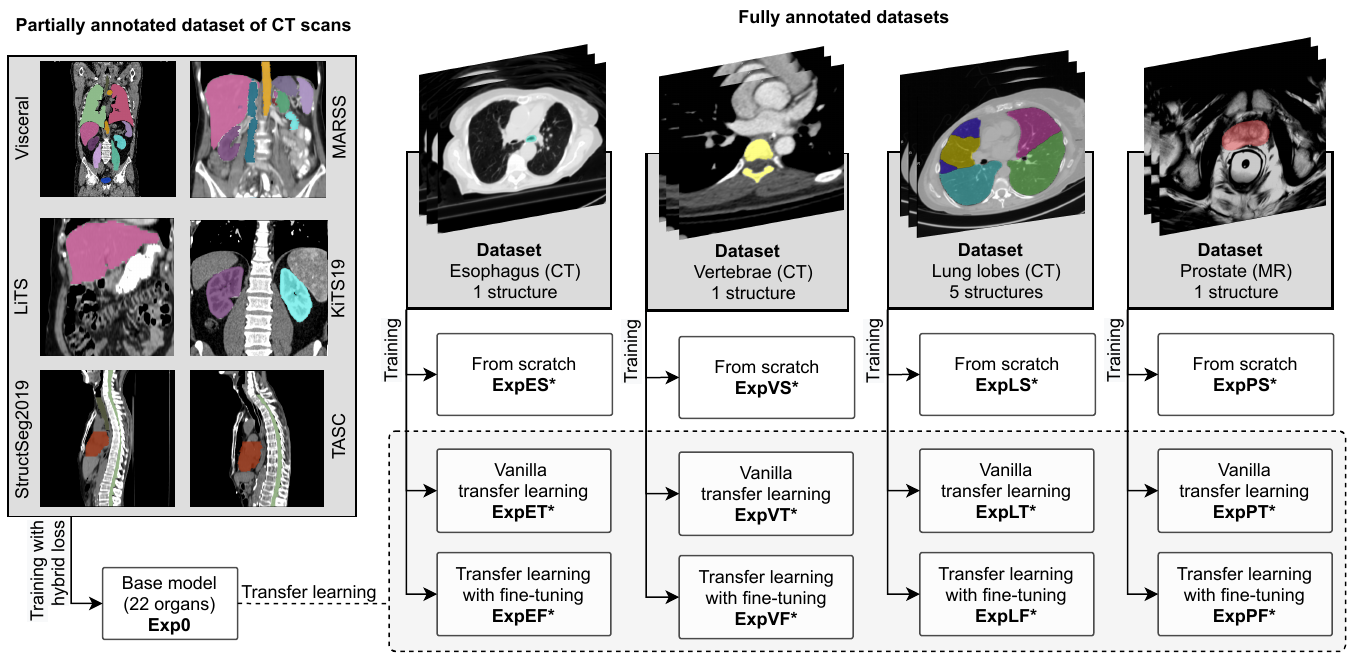}
    \caption{This paper has two types of datasets, the large sparsely annotated dataset--obtained by joining multiple publicly available datasets--and the fully annotated datasets (vertebrae/CT, esophagus/CT, lung lobes/CT, and prostate/MR). The large sparsely annotated dataset served to train Exp0, which learned from annotated regions only. The fully annotated datasets were used for the evaluation of three training strategies (scratch, vanilla transfer learning, and transfer learning with fine-tuning); each of these experiments was evaluated on different training set sizes. The weights of Exp0 initialized the weights of the networks that use transfer learning and continue training on the new fully annotated datasets.}
    \label{fig:diagram_proposal}
\end{figure*}

We used the remaining four structures (esophagus, vertebrae, lung lobes, and prostate) for transfer learning experiments with the generic base model (\Cref{fig:diagram_proposal}). The esophagus was annotated in three datasets already included in the training set, i.e., this new target structure was not part of the annotations in the training set, but the same CT scans were used to train the generic base model. This was not the case for the vertebrae and the lung lobes, which were annotated in scans that were not part of the training set. Finally, the prostate was annotated in MR scans rather than CT scans. For all four tasks, 10\% of the available scans were set aside for evaluation of the segmentation performance.

\section{Method}
The proposed strategy for training a generic base model using a sparsely annotated dataset does not require a specific network architecture. However, networks initialized with the weights of this base model will typically use the same or a very similar architecture since the learned weights are coupled to the size and order of the individual layers in the original architecture. We used the 3D U-Net~\citep{Cice16} architecture both for the generic base model and for all transfer learning experiments because this architecture is particularly popular for medical image segmentation tasks. However, to avoid potential issues with normalization layers that become too data-specific and thus hinder transfer learning, we opted to remove batch normalization from this architecture.
The 3D U-Net architecture used in this paper uses four resolutions, i.e., contains three pooling layers in the compression path. The number of filters in the convolutional layers starts at 32 filters and doubles after each pooling layer.

The 3D U-Net is a patch-based segmentation network. The models trained with CT images used an input patch size of 132$\times$132$\times$132 voxels. Since the network does not make use of padding in the convolutional layers, the size of the output patches is smaller, namely 44$\times$44$\times$44 voxels. The models trained with MR images from the PROMISE12 dataset used a different patch size of 108$\times$108$\times$108 voxels because the images had a substantially smaller field of view compared with the CT scans so that 132$\times$132$\times$132 voxel patches would have been often larger than the entire image. The corresponding output patch size was 20$\times$20$\times$20 voxels. Note that the network does not contain any fully-connected layers but only convolutions and pooling layers, which makes it possible to change the input patch size without affecting the model.

\subsection{Pre-processing}
All networks were trained with isotropically resampled images and reference segmentation masks. Images were resampled using cubic interpolation and reference segmentation masks using nearest neighbor interpolation. The combination of various datasets with images acquired in different institutions and for different purposes resulted in a dataset with a wide range of different image resolutions. For instance, the spacing between slices ranged from 0.5~mm (PancreasCT) to 5~mm (\gls{kits19}, \gls{lits} and \gls{cranialvault}). We resampled all images using cubic interpolation and all segmentation masks using nearest-neighbor interpolation to  1mm $\times$ 1mm $\times$ 1mm isotropic resolution.

Image intensities were normalized differently in CT and MR images. In CT images, the image values in Hounsfield Units (HU) were clipped to the range [-500,400], which roughly corresponds to the abdominal window and level settings used to view abdominal CT images. In MR images, we clipped the intensity values to the 5\% and 95\% percentiles of the image values (per image) and scaled this interval to the range [-500, 400] to match the range used for the CT data.

\begin{table*}[t]
    \caption{Data distribution per dataset for the four new tasks for the transfer learning experiments.}
    \label{tab:datasetdistribution}
    \renewcommand{\arraystretch}{1.2}
    \centering
    \footnotesize
    \begin{tabular}{llcccc}
        \hline
         \textbf{Segmentation} & \centering \multirow{2}{*}{\textbf{Data source}} & \multirow{2}{*}{\textbf{Modality}} & \multicolumn{3}{c}{\textbf{Number of scans}} \\[0.05cm] \cline{4-6}
         \centering\textbf{task}& & & \textbf{Training}& \textbf{Validation}& \textbf{Total} \\[0.1cm]
         \hline
         Esophagus & \gls{eligibson}, \gls{structseg}, and \gls{aapm} & CT & 158 & 18 & 176\\
         Vertebrae & \gls{csi} and \gls{verse19} & CT & 90 & 10 & 100 \\
         Lung lobes & COPDgene & CT &90 & 10 & 100 \\
         Prostate & PROMISE12 & MR & 45 & 5 & 50 \\[0.075cm] \hline
    \end{tabular}
\end{table*}

\subsection{Base model (Exp0)}
\label{sec:weightmap}

We refer to the generic base model that was trained with a dataset with sparse annotations of 22 structures as Experiment Zero (Exp0). This model was trained using patches evenly sampled from the training dataset with a stride of 30mm, a parameter determined through empirical experimentation. The final layer of the network is a softmax layer with 23 classes corresponding to 22 foreground structures and the background. The background contains all the structures that were not annotated in any of the images in the training set. Due to the sparse nature of the annotations, there are no images with annotations of all foreground structures, and hence no images in which the background voxels are known. For example, voxels that were not annotated as kidneys in the \Gls{kits19} images might be background voxels but might just as well belong to one of the other foreground classes that were not annotated in the \Gls{kits19} subset, e.g., the liver. To mitigate this problem, we trained the base model using a hybrid masked loss function composed of two terms, an average Dice score term and a cross-entropy term. These were computed only for the present foreground structures, ignoring classes that were not annotated in an image.

Given a binary flag $\delta_c \in \{0,1\}$ that indicates whether structure $c$ is one of the $N \in \{1,\dots,22\}$ annotated structures in the present image and a weight map $\omega_c$ (defined below), the hybrid masked loss function is defined as:
\begin{equation*}
    \mathcal{L} =
        \left( 1 - \frac{1}{N} \sum_{c=1}^{22} \delta_c \cdot \textrm{Dice}_c \right)
        +
        \left( - \sum_{c=1}^{22} \delta_c \cdot \omega_c \cdot \textrm{CCE}_c \right),
\end{equation*}

\noindent
where Dice$_c$ and CCE$_c$ correspond to the soft Dice score and the categorical cross-entropy for class $c$, respectively. Both loss components are never computed for the 23rd class, the background since background labels are unknown in any images. The factor $\delta_c$ ensures that the loss components are zero for structures that were not annotated in the image. Note that this does, in principle, not have any effect on the cross-entropy term since the cross-entropy for structure $c$ is already zero for voxels with another label.

The Dice score component is crucial for training with sparsely annotated data without explicit background labels. In each image, $N$ of the 22 structures are annotated, and the corresponding voxels' labels are thus known. For the remaining voxels, we only know that their labels are not among the $N$ annotated classes. The Dice score as a volume overlap measure penalizes false positive predictions and thus penalizes the classification of unlabeled voxels as one of the $N$ known structures. By presenting in consecutive training steps images from different subsets, i.e., with different structures annotated, the network can learn to recognize all 22 structures. The use of a softmax layer forces the network to resort to the background class for voxels that it does not recognize as any of the 22 structures; the probability of the background class will automatically increase when the network assigns low probabilities to all other classes.

The cross-entropy component is computed based on only the labeled voxels and thus only penalizes false-negative predictions. In combination with the Dice score component, it provides an additional penalty for incorrect classifications of the labeled voxels in a training image, which boosts the importance of the strong labels in the training data, i.e., the annotated foreground objects in each image. Because cross-entropy terms are sensitive to class imbalances, we also introduce a weight map $\omega_c$ for each class. The weight of each class corresponds to the inverse sampling probability across the entire training set \cite{Chen20a}.

The weights of the model were initialized using the Glorot uniform initializer \citep{Glor10}. To optimize the weights, we used the Adam optimizer \citep{King15} and trained three networks per experiment using different learning rates: \num{1e-4}, \num{1e-5}, and \num{5e-5}. The training was stopped once the mean dice score in the validation set did not improve for ten epochs. An epoch was defined as the full iteration of positive patches (containing annotations) from all CT images in the training set.

\subsection{Data augmentation}
During network training, 70\% of the samples in each epoch were subject to slight random transformations to augment the training data. We applied a combination of up to three of the following data augmentation techniques to the samples: 3D scaling, 3D rotation, Gaussian blurring, image intensity variation, and elastic deformations. Scaling was between -5\% and 5\%, rotation in up to two planes between -5$^\circ$ and 5$^\circ$ degrees, Gaussian blurring with sigma between 0.2 and 1.0, and image intensity variation between -20 and 20 HU, which was applied to the entire image. The elastic deformation method used a ten-control point grid on the sample where every control point was randomly shifted up to 5 voxels. Because the combination of elastic deformations and scaling or rotation frequently resulted in unrealistic images, we allowed only the combination with Gaussian blurring or image intensity variation.

Additionally, to combat the difference in image quality across scans from the various datasets, we randomly applied salt and pepper noise to 20\% of the voxels followed by Gaussian smoothing ($\sigma = 0.9$mm), which results in CT scans that look similar to scans acquired with lower radiation, i.e., lower mAs and kVp.

\subsection{Post-processing}
\label{sec:postproc}
The image was divided into non-overlapping patches corresponding to the network output to obtain segmentations of an entire image. The predictions for these patches were stitched together to form a complete segmentation mask and were thresholded at 0.5, assigning the background label if none of the foreground classes reached a probability above 0.5. Connected component analysis was used to remove all but the largest structure for each foreground class. Finally, the predicted segmentation masks were resampled to the image's original resolution.

\subsection{Transfer learning (\expe{XYZ})}

The effect of transfer learning when training a 3D U-Net for segmentation of a new structure was evaluated for four new segmentation tasks (\Cref{tab:datasetdistribution}). For each task, we compared three training strategies: (1) training from scratch, where the weights of the network were randomly initialized, and all network weights were updated during training; (2) vanilla transfer learning, where the weights of the network were initialized with the weights of the trained base model (Exp0), and all networks weights are updated during training; and (3) transfer learning in combination with fine-tuning, where the base model was used to initialize the network, followed by training only the last three layers for the new task while the other layers remained fixed. The intention was to prevent a form of catastrophic forgetting where useful low-level filters in the first layers that were inherited from the base model might be forgotten when switching abruptly to a new task. By first adjusting only the last few layers to the new task and then fine-tuning the entire network, the network might profit more from transfer learning in a second step. In each step, the network was trained until convergence.

The output layer of the base model has 23 channels, corresponding to a background class and 22 foreground classes. New tasks will usually have fewer classes, e.g., three of our example tasks are binary segmentation problems with only one foreground class, and the lung lobe segmentation task has five foreground classes. When initializing a new model with the weights of the base model, we retain the background class and reduce the number of foreground classes to the required number of foreground classes by dropping channels from the output layer.

The new models were trained in the same way as the base model with the exception of the loss function, which did not use the class presence factor $\delta_c$ and included the background class in the cross entropy computation (i.e., an unmodified weighted cross-entropy term was used in combination with an unmodified soft Dice score term). In addition, we extracted patches with a stride of 10mm from the images because there were fewer images in the training sets than in the base model training set which used a stride of 30mm.

\begin{table*}[ht!]
    \scriptsize
    \centering
    \renewcommand{\arraystretch}{1.2}
    \caption{List of experiment IDs for the four additional segmentation tasks (vertebrae, esophagus, lung lobes, and prostate), training strategy (scratch, vanilla transfer learning, and transfer learning with fine-tuning), and training set size ($Z=[10,20,30,40,50, $ and full$]$  CT scans). For instance, the experiment \expe{VT40} trained on 40 CT scans using vanilla transfer learning to segment the vertebrae. The results show the mean Dice score and standard deviation.}
    \begin{tabular}{p{3.1cm}cccccc}
    \hline\\[-0.3cm]
    & \multicolumn{2}{c}{\multirow{2}{*}{\textbf{Training from scratch}}} & \multicolumn{2}{c}{\multirow{2}{*}{\textbf{Vanilla transfer learning}}} & \multicolumn{2}{c}{\textbf{Transfer learning}}\\[-0.115cm]
    & & & & & \multicolumn{2}{c}{\textbf{with fine-tuning step}}\\[0.05cm]
    & Experiment & Dice score & Experiment & Dice score & Experiment & Dice score\\\hline%
    & \expe{ES10} & 0.459 $\pm$ 0.245 & \expe{ET10} & 0.548 $\pm$ 0.214 & \expe{EF10} & 0.588 $\pm$ 0.193 \\
     \textbf{Esophagus} & \expe{ES20} & 0.579 $\pm$ 0.194 & \expe{ET20} & 0.617 $\pm$ 0.201 & \expe{EF20} & 0.641 $\pm$ 0.197 \\
    & \expe{ES30} & 0.590 $\pm$ 0.204 & \expe{ET30} & 0.632 $\pm$ 0.197 & \expe{EF30} & 0.652 $\pm$ 0.194 \\
    Modality:~CT & \expe{ES40} & 0.624 $\pm$ 0.195 & \expe{ET40} & 0.659 $\pm$ 0.195 & \expe{EF40} & 0.666 $\pm$ 0.197 \\
    Structures:~1 & \expe{ES50} & 0.627 $\pm$ 0.213 & \expe{ET50} & 0.653 $\pm$ 0.196 & \expe{EF50} & 0.668 $\pm$ 0.197 \\
    & \expe{ES158} & 0.694 $\pm$ 0.196 & \expe{ET158} & 0.684 $\pm$ 0.206 & \expe{EF158} & 0.696 $\pm$ 0.206 \\ \hline%
    & \expe{VS10} & 0.920 $\pm$ 0.036 & \expe{VT10} & 0.926 $\pm$ 0.031 & \expe{VF10} & 0.929 $\pm$ 0.028 \\
    \textbf{Vertebrae} & \expe{VS20} & 0.931 $\pm$ 0.030 & \expe{VT20} & 0.935 $\pm$ 0.029 & \expe{VF20} & 0.928 $\pm$ 0.029 \\
    & \expe{VS30} & 0.939 $\pm$ 0.023 & \expe{VT30} & 0.942 $\pm$ 0.025 & \expe{VF30} & 0.939 $\pm$ 0.025 \\
    Modality:~CT~(unseen) & \expe{VS40} & 0.943 $\pm$ 0.022 & \expe{VT40} & 0.944 $\pm$ 0.025 & \expe{VF40} & 0.943 $\pm$ 0.026 \\
    Structures:~1 & \expe{VS50} & 0.942 $\pm$ 0.024 & \expe{VT50} & 0.952 $\pm$ 0.020 & \expe{VF50} & 0.949 $\pm$ 0.023 \\
    & \expe{VS90} & 0.956 $\pm$ 0.020 & \expe{VT90} & 0.955 $\pm$ 0.022 & \expe{VF90} & 0.955 $\pm$ 0.016 \\ \hline%
    & \expe{LS10} & 0.917 $\pm$ 0.027 & \expe{LT10} & 0.930 $\pm$ 0.028 & \expe{LF10} & 0.941 $\pm$ 0.030 \\
    \textbf{Lung lobes} & \expe{LS20} & 0.939 $\pm$ 0.035 & \expe{LT20} & 0.948 $\pm$ 0.026 & \expe{LF20} & 0.950 $\pm$ 0.025 \\
    & \expe{LS30} & 0.951 $\pm$ 0.024 & \expe{LT30} & 0.955 $\pm$ 0.022 & \expe{LF30} & 0.956 $\pm$ 0.023 \\
    Modality:~CT~(unseen) & \expe{LS40} & 0.959 $\pm$ 0.016 & \expe{LT40} & 0.961 $\pm$ 0.016 & \expe{LF40} & 0.961 $\pm$ 0.018 \\
    Structures:~5 & \expe{LS50} & 0.961 $\pm$ 0.015 & \expe{LT50} & 0.965 $\pm$ 0.014 & \expe{LF50} & 0.964 $\pm$ 0.013 \\
    & \expe{LS90} & 0.969 $\pm$ 0.010 & \expe{LT90} & 0.970 $\pm$ 0.011 & \expe{LT90} & 0.969 $\pm$ 0.013 \\ \hline%
    & \expe{PS10} & 0.816 $\pm$ 0.020 & \expe{PT10} & 0.813 $\pm$ 0.014 & \expe{PF10} & 0.818 $\pm$ 0.019 \\
    \textbf{Prostate} & \expe{PS20} & 0.819 $\pm$ 0.062 & \expe{PT20} & 0.851 $\pm$ 0.032 & \expe{PF20} & 0.862 $\pm$ 0.018 \\
    & \expe{PS30} & 0.844 $\pm$ 0.053 & \expe{PT30} & 0.854 $\pm$ 0.018 & \expe{PF30} & 0.870 $\pm$ 0.015 \\
    Modality:~MR (unseen) & \expe{PS40} & 0.852 $\pm$ 0.070 & \expe{PT40} & 0.861 $\pm$ 0.034 & \expe{PF40} & 0.869 $\pm$ 0.016 \\
    Structures:~1 & \expe{PS50} & -- & \expe{PT50} & -- & \expe{PF50} & -- \\
    & \expe{PS45} & 0.852 $\pm$ 0.031 & \expe{PT45} & 0.882 $\pm$ 0.010 & \expe{PF45} & 0.884 $\pm$ 0.012 \\ \hline
    \end{tabular}
    \label{tab:experiments}
\end{table*}

\begin{figure*}[ht!]
    \centering
    \begin{subfigure}[b]{0.49\textwidth}
    \includegraphics[width=\textwidth]{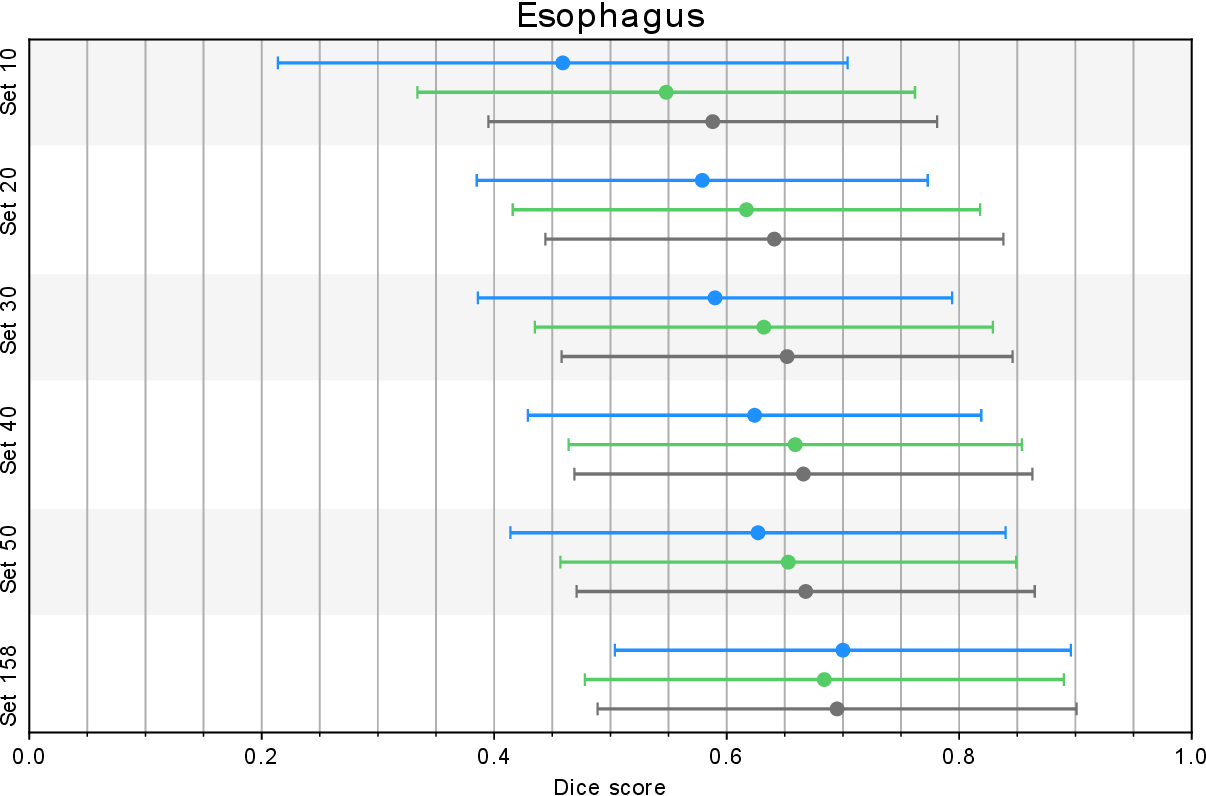}
  \end{subfigure}\hfill%
  \begin{subfigure}[b]{0.49\textwidth}
    \includegraphics[width=\textwidth]{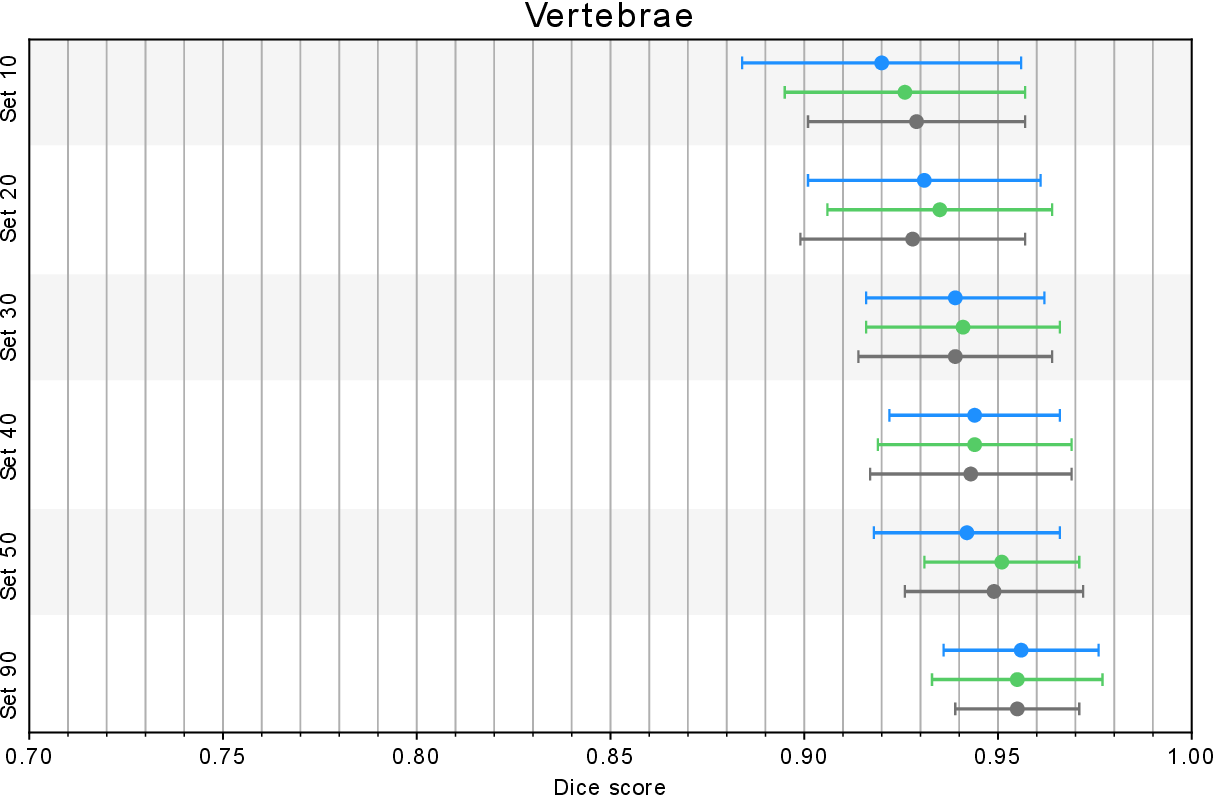}
  \end{subfigure}\hfill%
  \vspace{0.3cm}
  \begin{subfigure}[b]{0.49\textwidth}
    \includegraphics[width=\textwidth]{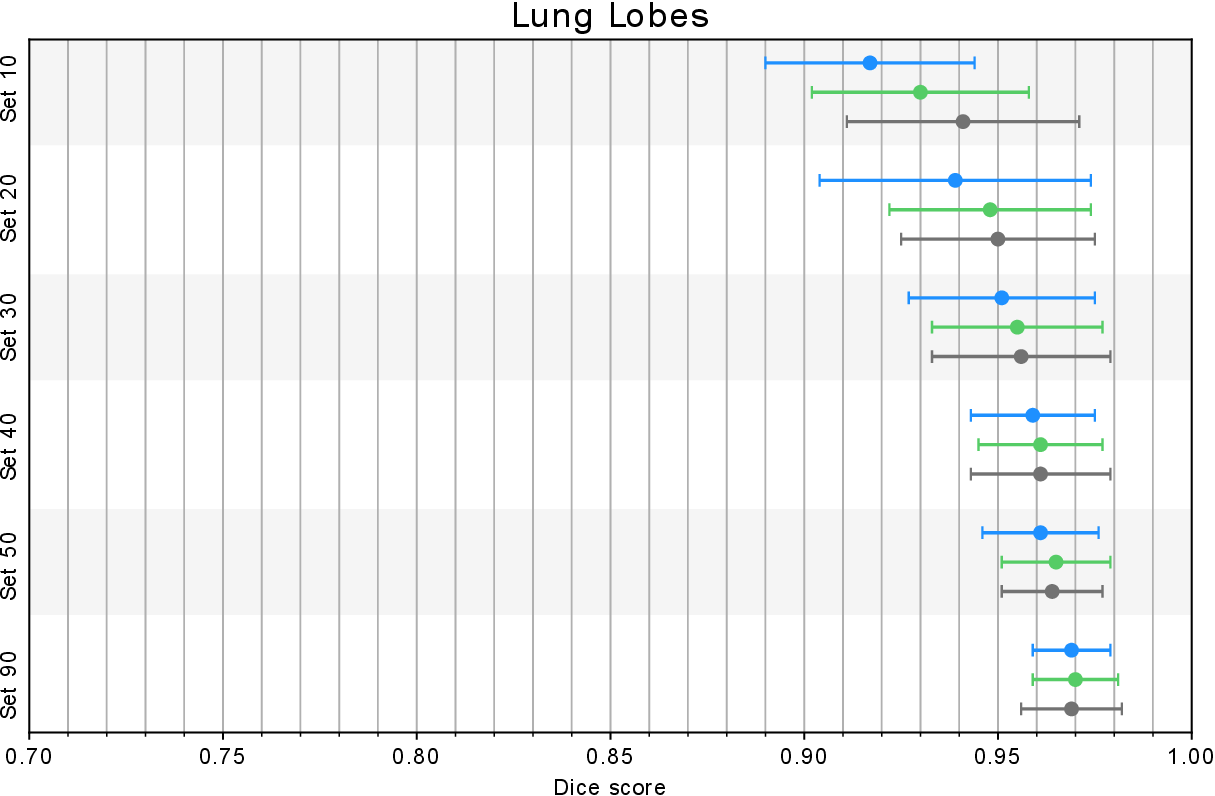}
  \end{subfigure}\hfill%
  \begin{subfigure}[b]{0.49\textwidth}
    \includegraphics[width=\textwidth]{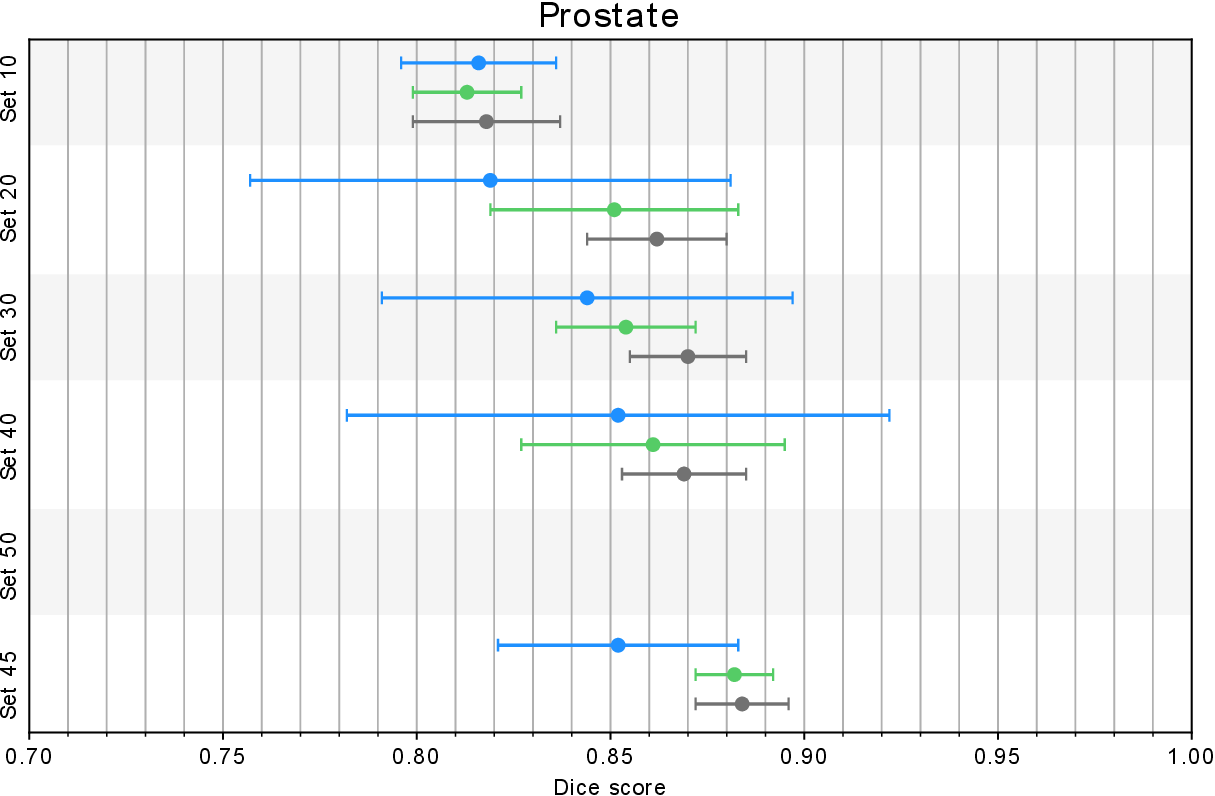}
  \end{subfigure}\hfill%
  \vspace{0.05cm}
  \begin{subfigure}[b]{0.8\textwidth}
    \includegraphics[width=\textwidth]{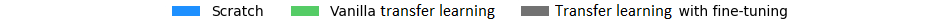}
  \end{subfigure}
    \caption{Bar plots of Dice score obtained by the experiments reported in \Cref{tab:experiments} respectively. Note that the Dice score ranges from 0.7 to 1.0 in the subfigures for better visualization, except for the esophagus.}
    \label{fig:resultsmeandice}
\end{figure*}

\section{Results}
We performed experiments with the base model and with models trained for four new segmentation tasks, where these models were either trained from scratch or initialized with the base model. The new models were trained with different training set sizes by randomly selecting a subset of the available training data to simulate limited training data.
In the following, we refer to the generic base model as Exp0 and use the following naming convention for the experiments of the four new tasks: \expe{XYZ}, where X represents the segmentation task (E=esophagus, V=vertebrae, L=lung lobes, and P=prostate), Y represents the training strategy (S=scratch, T=vanilla transfer learning, F=transfer learning with fine-tuning step), and Z represents the training set size (10, 20, 30, 40, 50, and all the images available in the dataset).

\subsection{Base model (Exp0)}
The base model was trained with a sparsely annotated dataset and evaluated on a randomly selected subset of this dataset, which was not used for training. Note that there was no separate test set but that the performance on the validation set is reported. We chose not to reserve a test set for evaluating the base model because the transfer learning experiments are the focus of this paper, and the base model's performance is only of secondary interest. Since the base model was evaluated with sparsely annotated data, we ignored structures without annotations and calculated the Dice volume overlap score for each image between the available reference segmentations and the automatic segmentation results for these structures. The scores were then averaged across the dataset, resulting in an average Dice score of 0.725 $\pm$ 0.195 for segmenting 22 structures across 54 images. The highest scores were achieved for large structures like the lungs (0.967 and 0.966 for the right and left lungs, respectively), while small and irregular structures like the portal and splenic veins (0.349) achieved the lowest scores.

\subsection{Transfer learning from Exp0}
To evaluate whether transfer learning from the generic base model (Exp0) is beneficial, we trained segmentation networks for four additional tasks using fully annotated rather than sparsely annotated datasets (\Cref{tab:datasetdistribution}). The performance of the models when trained with the full datasets and smaller subsets of the data are listed in \Cref{tab:experiments} and visualized in \Cref{fig:resultsmeandice}. \Cref{fig:predictionsset30esophagus,fig:predictionsset30prostate,fig:predictionsset30lunglobes,fig:predictionsset30vertebrae} show examples of segmentations; the green and red regions represent the annotations and the predictions, respectively, both with transparency to visualize the overlap (dark green) between regions.

The segmentation models generally performed better when trained with larger training sets, regardless of the training strategy. When training from scratch, the increase in performance when training with more data was largest in the low data regime. For example, the Dice score for the challenging esophagus segmentation task increased from 0.459 to 0.579 (diff. +0.120) when training with 20 instead of 10 scans, but adding ten additional scans and training with 30 scans in total only resulted in a further increase of 0.11 in Dice score.
For the pulmonary lobe segmentation task, where large parts of the object are well recognizable but where the boundaries are challenging to delineate precisely, ten scans were sufficient to reach a Dice score of 0.917. Adding more scans resulted in a steady increase in performance, reaching 0.969 when training with 90 scans.

Transfer learning by initializing the network weights with the weights of the base model resulted, in most cases, in a better segmentation performance, both vanilla and with the fine-tuning step. The impact differed per task and training set size. Models that were trained with a small dataset and reached low segmentation performance generally profited more from transfer learning, such as the esophagus segmentation task where the model trained with only ten scans improved from 0.459 to 0.548 (diff. +0.089) when using vanilla transfer learning and to 0.588 when using transfer learning with fine-tuning step. On the other hand, lung lobes and vertebra segmentation models trained with ten scans already reached a Dice score of more than 0.900, and transfer learning had a limited impact on the performance of these models. Overall, we observed that transfer learning did usually not contribute anymore to the increase in performance when approximately 30 training scans were available. At the same time, transfer learning usually did not hurt the performance.

\subsection{Cross-modality transfer learning}


To evaluate whether transfer learning from the base model (Exp0, trained on CT scans) is beneficial for other modalities, we trained networks using MR scans (prostate segmentation task) as input instead of CT scans.
Although all the training strategies trained on ten images obtained similar results, both transfer learning strategies obtained higher results than experiments trained from scratch when training on 20 (and more) images.
When training with 20 images, the vanilla transfer learning experiment obtained 0.851 Dice, while the experiment trained from scratch obtained 0.816 (diff: +0.032); the fine-tuning experiment obtained 0.862 (diff with scratch +0.062). As with the experiments with CT scans (esophagus, lung lobes, and vertebrae segmentation tasks), the performance increases when adding more images to the training set regardless of the training strategy.
We observed that the difference among training strategies gets smaller with more training data for CT data; this was different for the prostate segmentation task, where the transfer learning experiments kept a steady difference with the experiments trained from scratch while adding more training data.
For instance, the difference in Dice score between vanilla transfer learning and scratch experiments ranged from 0.010 to 0.032 (average diff: 0.020) when the training set size increased from 10 to 45 MR scans. While the difference in Dice score between transfer learning with fine-tuning and scratch experiments ranged from 0.026 to 0.043 (average diff: 0.029) when increasing the training set size from 10 to 45 MR scans.

\subsection{Transfer learning with fine-tuning step from Exp0}
Large network weight changes may happen when a pretrained network is re-trained on a different task.
We conducted experiments where the pretrained network gradually adapted to the new task by allowing weight changes to a certain number of layers.
Subsequently, changes to all the network weights are allowed for further specialization; we refer to this procedure as transfer learning with fine-tuning step.

Overall, transfer learning with fine-tuning obtained higher results than the other two training strategies (vanilla transfer learning and scratch).
For instance, the experiment to segment the esophagus trained on 10 CT scans using transfer learning with fine-tuning (\expe{EF10}) obtained 0.588 Dice score, +0.129 than the experiments trained from scratch \expe{ES10}. Adding more images to the training set gradually reduces the performance difference from 0.129 to 0.002 (average diff: 0.056). 
The vertebrae segmentation tasks slightly increased the performance of the experiments trained from scratch from 0.920 to 0.929 (diff: +0.009) when training on 10 CT scans; adding more CT scans made the experiments trained from scratch slightly better than transfer learning.
Similarly, the lung lobe segmentation task showed minor improvements, where fine-tuning with 10 CT scans obtained 0.941 Dice, which is only +0.024 higher than the experiment trained from scratch.

When comparing the transfer learning strategies, we observe that transfer learning with fine-tuning gets higher results than vanilla transfer learning.
The esophagus was the most benefited segmentation task with the transfer learning with fine-tuning strategy.
For instance, for a training set of 10 CT scans, the performance increased in +0.040 Dice score when using fine-tuning compared to vanilla transfer learning.
Fine-tuning got slightly higher results for the lung lobes than vanilla transfer learning when training with up to 30 images, while more images made vanilla transfer learning slightly higher.
For the vertebrae experiments, the vanilla transfer learning was slightly better than fine-tuning; the difference in performance was between -0.0066 to +0.0026 (average diff: -0.0017) for the experiments with different training set sizes.

\subsection{Vanilla transfer learning from a simpler model}
To evaluate whether transfer learning from Exp0 (trained on a large sparsely annotated dataset) is more beneficial than a simpler model, we compared the results of transfer learning from a simpler model to transfer learning from Exp0.
We conducted experiments using vanilla transfer learning from:

\subsubsection{Vertebrae to other segmentation tasks}
We picked one of the models of the new segmentation tasks trained from scratch with all the images available for further analysis.
The weights of the model \expe{VS90} (vertebrae trained from scratch using 90 CT scans) initialized networks to perform vanilla transfer learning on the esophagus, lung lobes, and prostate datasets (see \Cref{tab:vanillatlvertebrae2others}).
The esophagus experiment (initialized with the simpler model \expe{VS90}) trained on 10 CT scans obtained 0.475 Dice, while \expe{ET10} (initialized with Exp0) obtained 0.548 Dice (diff: +0.073).
The difference in all the training set sizes for the esophagus ranged from -0.004 to 0.073 (average diff among five experiments: 0.036).
Similarly, for the lung lobe segmentation task, the difference in performance among training set sizes ranges from 0.001 to 0.022 (average diff among five experiments: 0.009).
In contrast, the prostate segmentation task obtained differences in performance ranging from -0.02 to 0.019 (average diff among four experiments: -0.002).

\begin{table*}[t]
    \caption{Results of vanilla transfer learning from a simpler model (\expe{VS90}) to the esophagus, lung lobes, and prostate segmentation tasks.}
    \label{tab:vanillatlvertebrae2others}
    \renewcommand{\arraystretch}{1.2}
    \centering
    \scriptsize
    \begin{tabular}{lcccccp{2cm}}
        \hline
         \textbf{Dataset} & \textbf{Set: 10 images} & \textbf{Set: 20 images} & \textbf{Set: 30 images} & \textbf{Set: 40 images} & \textbf{Set: 50 images} & \textbf{Set: all images available}\\ \hline
         Esophagus & 0.475 $\pm$ 0.243 & 0.550 $\pm$ 0.254 & 0.617 $\pm$ 0.199 & 0.616 $\pm$ 0.218 & 0.633 $\pm$ 0.199 & 0.688 $\pm$ 0.209\\
         Lung lobes &  0.908 $\pm$ 0.028 & 0.934 $\pm$ 0.029 & 0.946 $\pm$ 0.029 & 0.956 $\pm$ 0.017 & 0.964 $\pm$ 0.014 & 0.968 $\pm$ 0.016\\
         Prostate & 0.794 $\pm$ 0.043 & 0.844 $\pm$ 0.035 & 0.870 $\pm$ 0.017 & 0.881 $\pm$ 0.026 & -- & 0.882 $\pm$ 0.017\\ \hline
    \end{tabular}
\end{table*}

\begin{figure*}[!ht]
\rotatebox{90}{\hspace{0.65cm}Esophagus 1}
    \begin{subfigure}[b]{0.24\textwidth}
    \includegraphics[width=\textwidth]{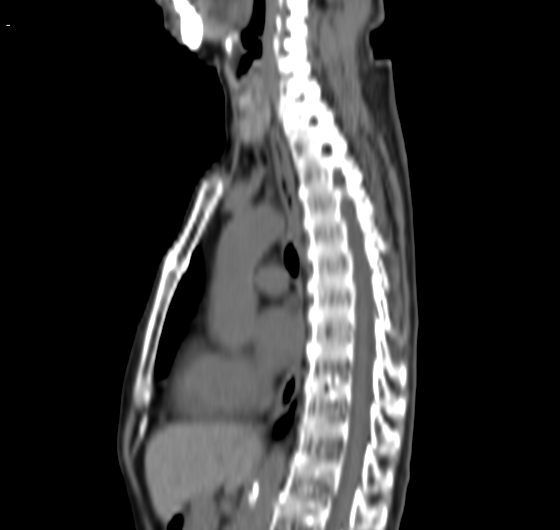}
  \end{subfigure}\hfill%
  \begin{subfigure}[b]{0.24\textwidth}
    \includegraphics[width=\textwidth]{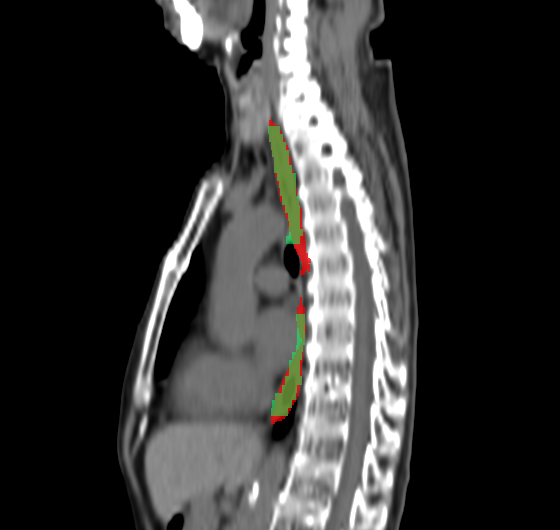}
  \end{subfigure}\hfill%
  \begin{subfigure}[b]{0.24\textwidth}
    \includegraphics[width=\textwidth]{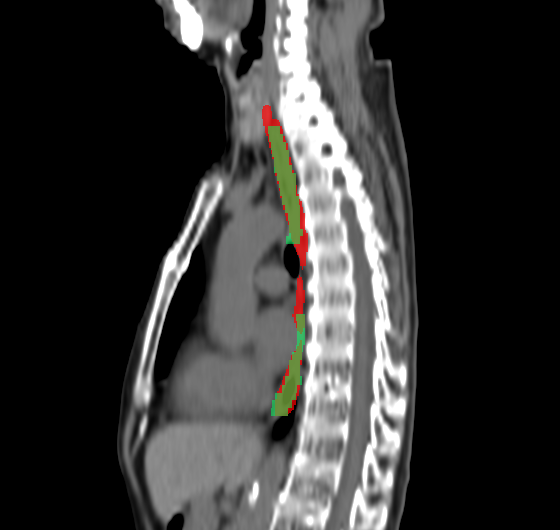}
  \end{subfigure}\hfill%
  \begin{subfigure}[b]{0.24\textwidth}
    \includegraphics[width=\textwidth]{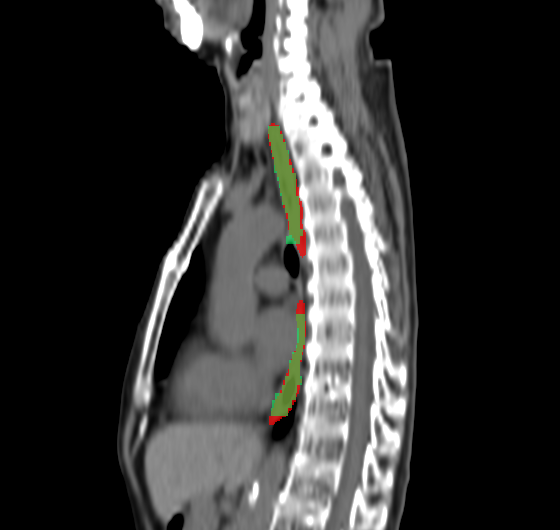}
  \end{subfigure}\hfill%
  
  \rotatebox{90}{\hspace{1.1cm}Esophagus 2}
  \begin{subfigure}[b]{0.24\textwidth}
    \includegraphics[width=\textwidth]{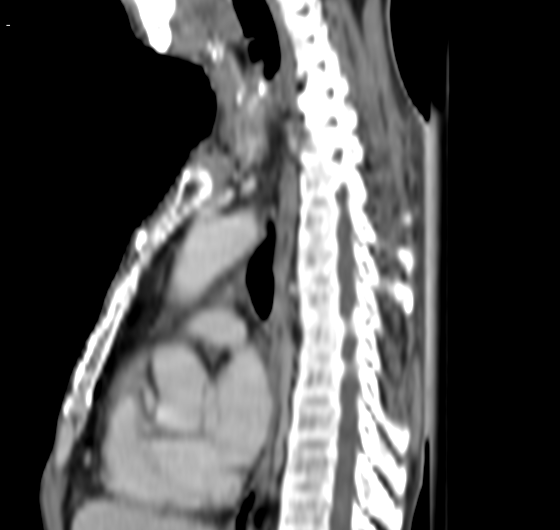}
    \caption{}
  \end{subfigure}\hfill%
  \begin{subfigure}[b]{0.24\textwidth}
    \includegraphics[width=\textwidth]{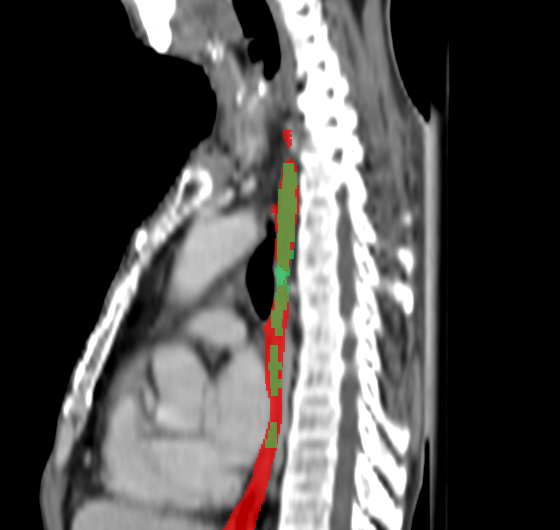}
    \caption{}
  \end{subfigure}\hfill%
  \begin{subfigure}[b]{0.24\textwidth}
    \includegraphics[width=\textwidth]{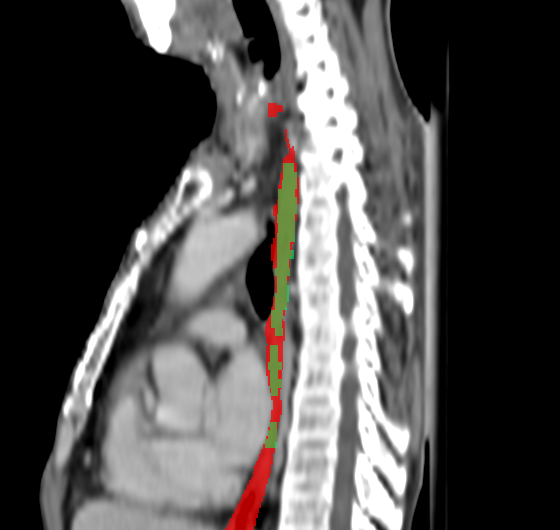}
    \caption{}
  \end{subfigure}\hfill%
  \begin{subfigure}[b]{0.24\textwidth}
    \includegraphics[width=\textwidth]{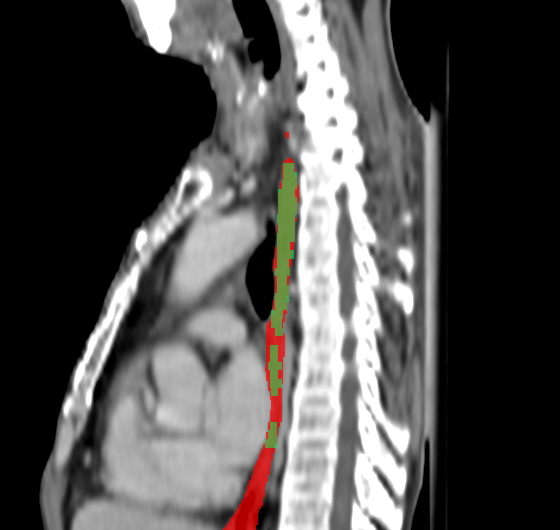}
    \caption{}
  \end{subfigure}\hfill%
  \caption{Predictions of the experiments of the esophagus segmentation task trained on 30 CT scans. (a) Shows the original slice, and the training strategies (b) scratch \expe{ES30}, (c) vanilla transfer learning \expe{ET30}, and (d) transfer learning with fine-tuning \expe{EF30}.}
  \label{fig:predictionsset30esophagus}
\end{figure*}

\begin{figure*}[!ht]
  \rotatebox{90}{\hspace{0.65cm}Vertebrae 1}
  \begin{subfigure}[b]{0.24\textwidth}
    \includegraphics[width=\textwidth]{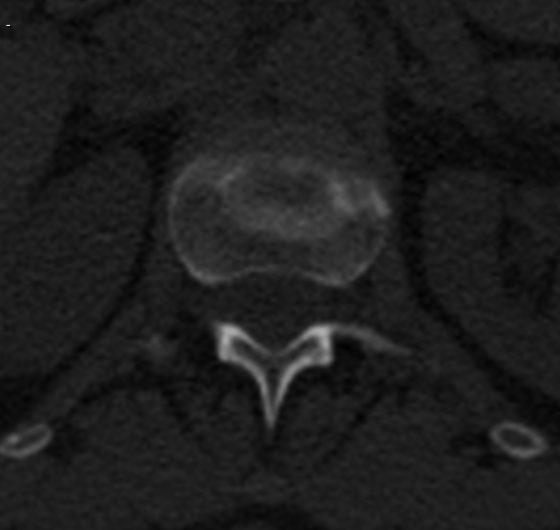}
  \end{subfigure}\hfill%
  \begin{subfigure}[b]{0.24\textwidth}
    \includegraphics[width=\textwidth]{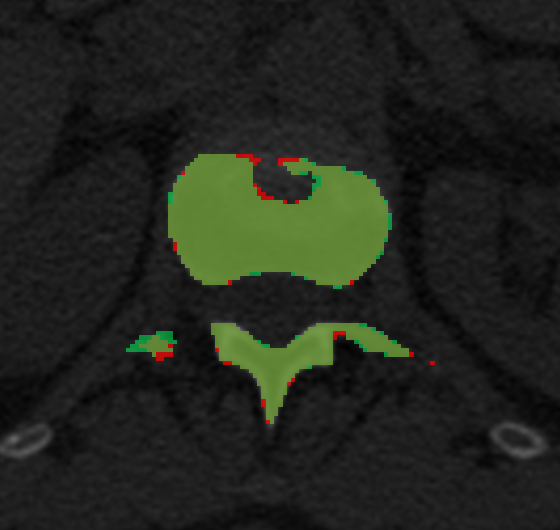}
  \end{subfigure}\hfill%
  \begin{subfigure}[b]{0.24\textwidth}
    \includegraphics[width=\textwidth]{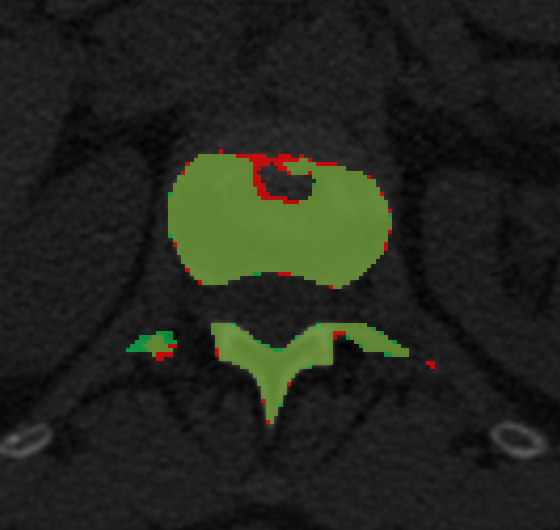}
  \end{subfigure}\hfill%
  \begin{subfigure}[b]{0.24\textwidth}
    \includegraphics[width=\textwidth]{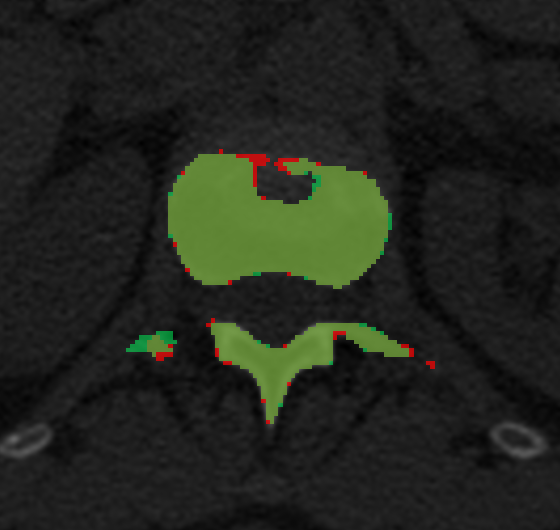}
  \end{subfigure}\hfill%
  
  \rotatebox{90}{\hspace{1.1cm}Vertebrae 2}
  \begin{subfigure}[b]{0.24\textwidth}
    \includegraphics[width=\textwidth]{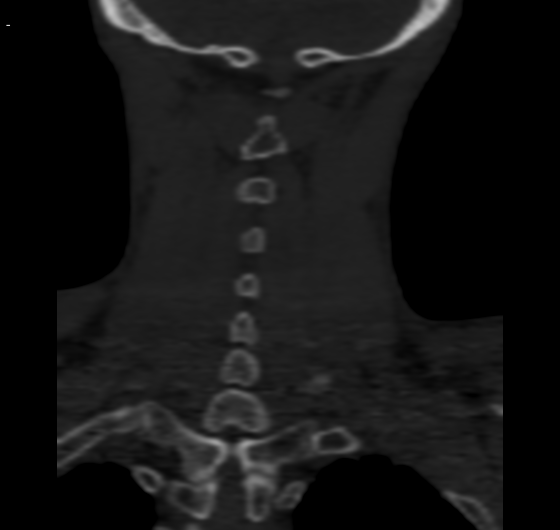}
    \caption{}
  \end{subfigure}\hfill%
  \begin{subfigure}[b]{0.24\textwidth}
    \includegraphics[width=\textwidth]{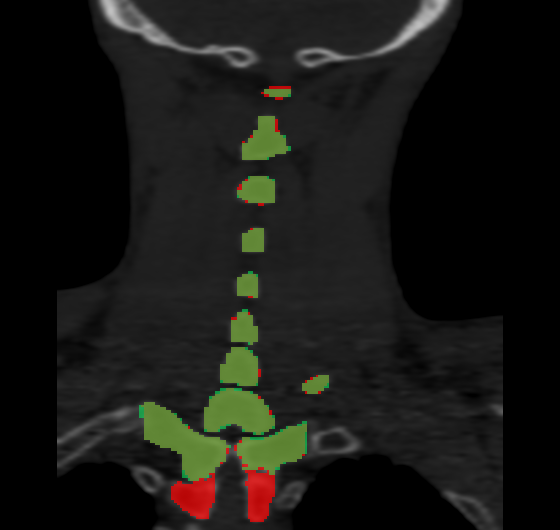}
    \caption{}
  \end{subfigure}\hfill%
  \begin{subfigure}[b]{0.24\textwidth}
    \includegraphics[width=\textwidth]{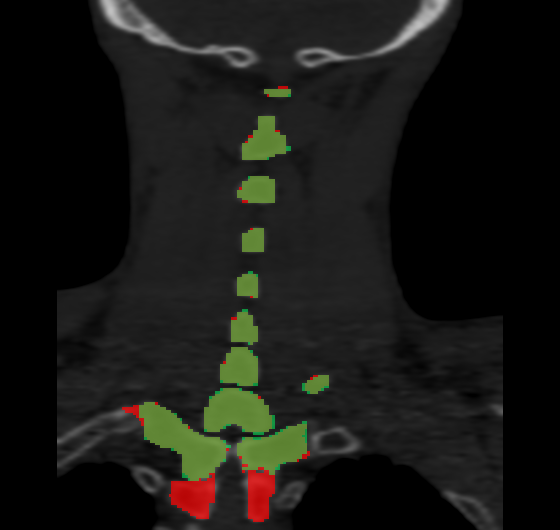}
    \caption{}
  \end{subfigure}\hfill%
  \begin{subfigure}[b]{0.24\textwidth}
    \includegraphics[width=\textwidth]{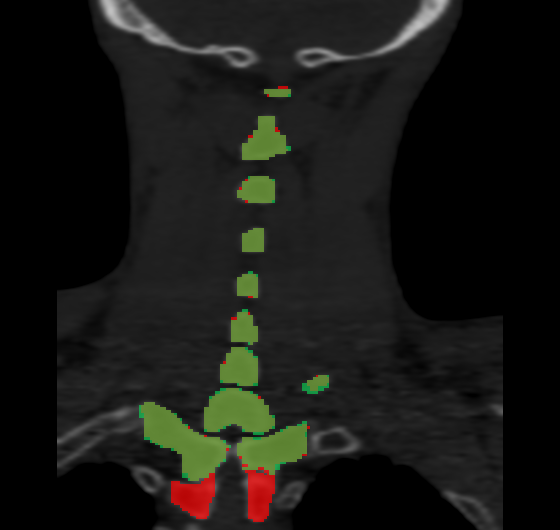}
    \caption{}
  \end{subfigure}\hfill%
  \caption{Predictions of the experiments of the vertebrae segmentation task trained on 30 CT scans. (a) Shows the original slice, and the training strategies (b) scratch \expe{VS30}, (c) vanilla transfer learning \expe{VT30}, and (d) transfer learning with fine-tuning \expe{VF30}.}
  \label{fig:predictionsset30vertebrae}
\end{figure*}

\begin{figure*}[!ht]
  \rotatebox{90}{\hspace{0.25cm}Prostate: Case 1}
  \begin{subfigure}[b]{0.24\textwidth}
    \includegraphics[width=\textwidth]{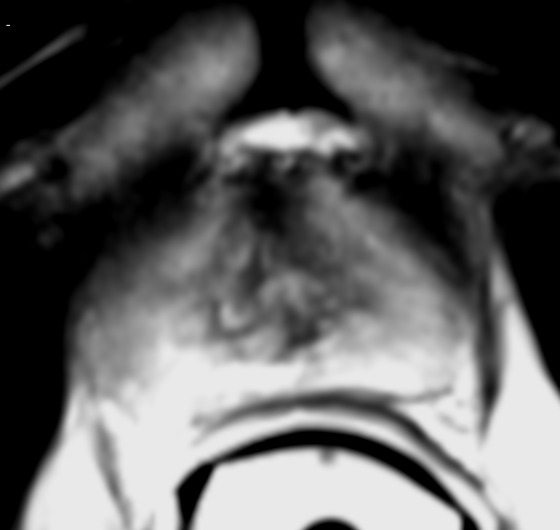}
  \end{subfigure}\hfill%
  \begin{subfigure}[b]{0.24\textwidth}
    \includegraphics[width=\textwidth]{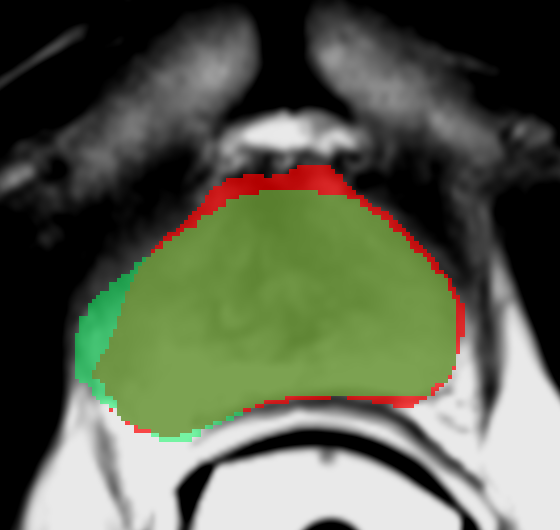}
  \end{subfigure}\hfill%
  \begin{subfigure}[b]{0.24\textwidth}
    \includegraphics[width=\textwidth]{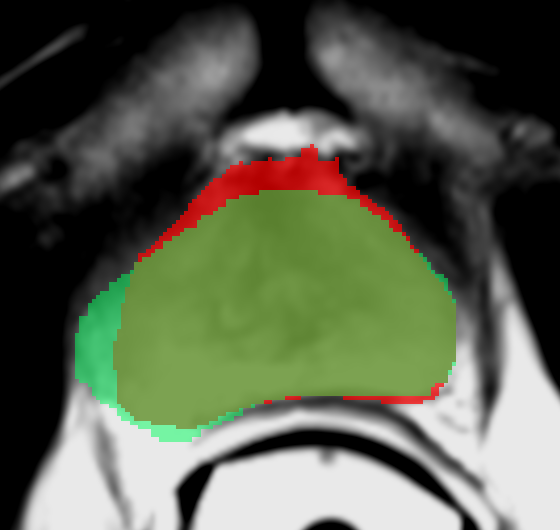}
  \end{subfigure}\hfill%
  \begin{subfigure}[b]{0.24\textwidth}
    \includegraphics[width=\textwidth]{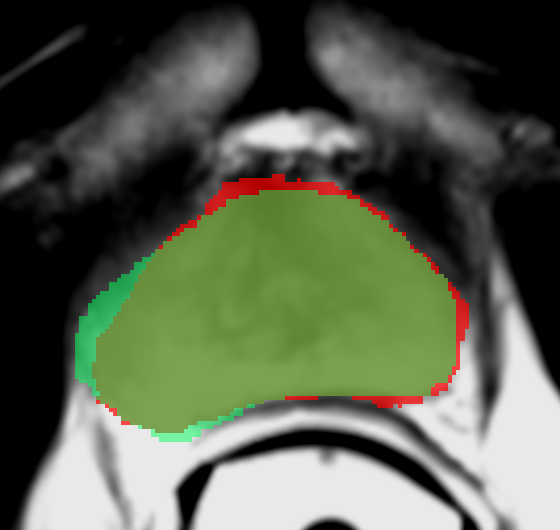}
  \end{subfigure}\hfill%
  
  \rotatebox{90}{\hspace{0.6cm}Prostate: Case 2}
  \begin{subfigure}[b]{0.24\textwidth}
    \includegraphics[width=\textwidth]{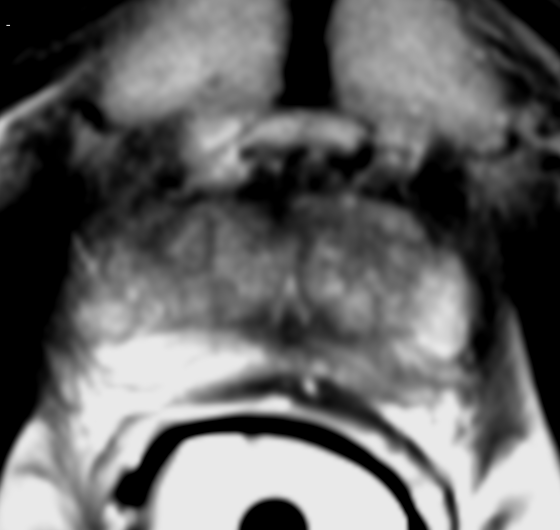}
    \caption{}
  \end{subfigure}\hfill%
  \begin{subfigure}[b]{0.24\textwidth}
    \includegraphics[width=\textwidth]{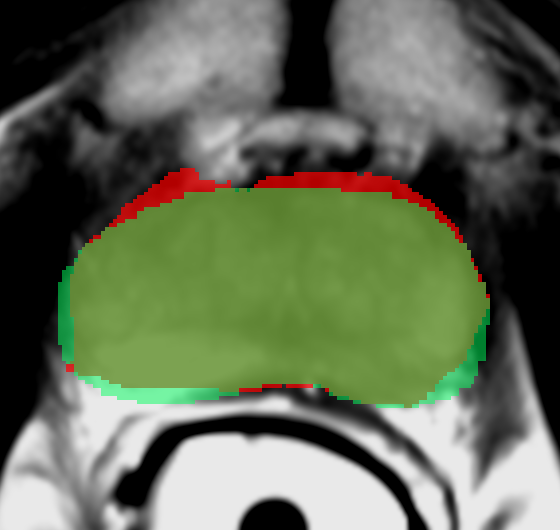}
    \caption{}
  \end{subfigure}\hfill%
  \begin{subfigure}[b]{0.24\textwidth}
    \includegraphics[width=\textwidth]{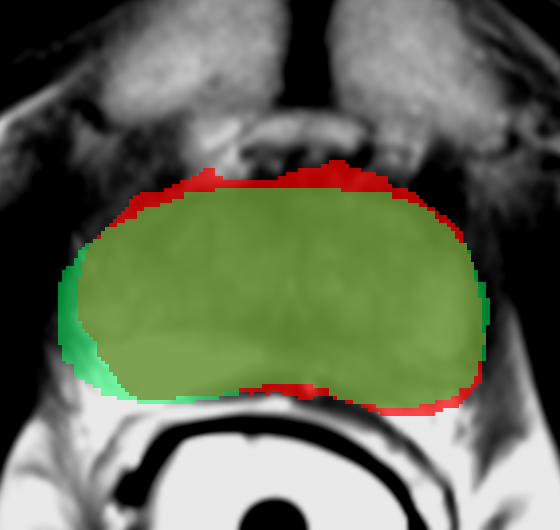}
    \caption{}
  \end{subfigure}\hfill%
  \begin{subfigure}[b]{0.24\textwidth}
    \includegraphics[width=\textwidth]{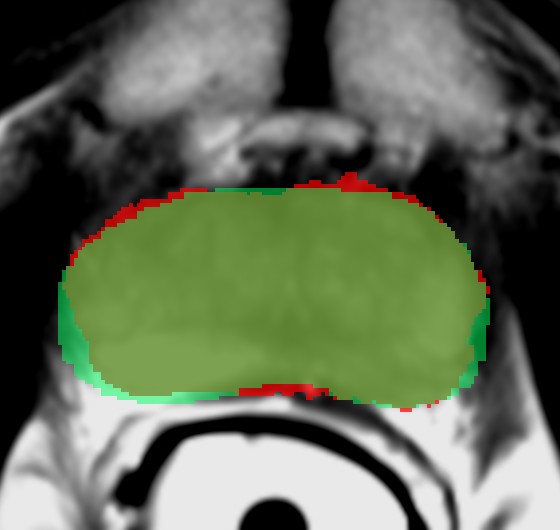}
    \caption{}
  \end{subfigure}\hfill%
  \caption{Predictions of the experiments of the prostate segmentation task trained on 30 MR scans. (a) Shows the original slice, and the training strategies (b) scratch \expe{PS30}, (c) vanilla transfer learning \expe{PT30}, and (d) transfer learning with fine-tuning \expe{PF30}.}
  \label{fig:predictionsset30prostate}
\end{figure*}

\begin{figure*}[!ht]
  \rotatebox{90}{\hspace{0.15cm}Lung lobes: Case 1}
  \begin{subfigure}[b]{0.24\textwidth}
    \includegraphics[width=\textwidth]{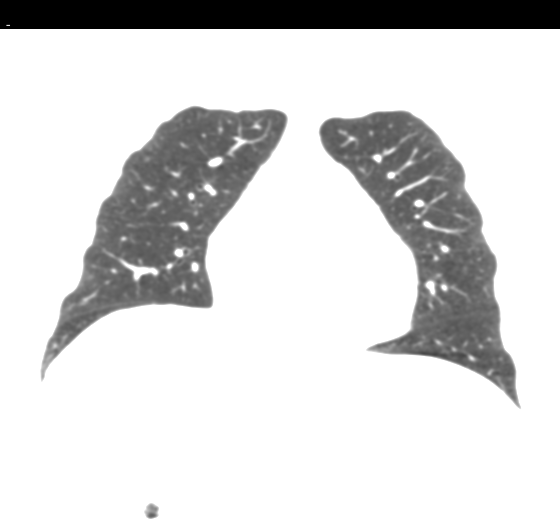}
  \end{subfigure}\hfill%
  \begin{subfigure}[b]{0.24\textwidth}
    \includegraphics[width=\textwidth]{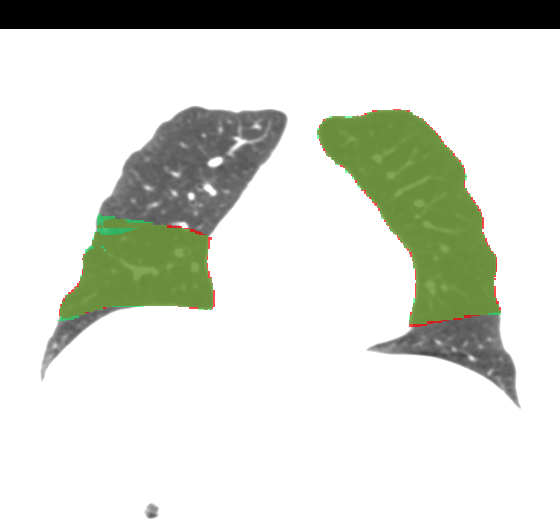}
  \end{subfigure}\hfill%
  \begin{subfigure}[b]{0.24\textwidth}
    \includegraphics[width=\textwidth]{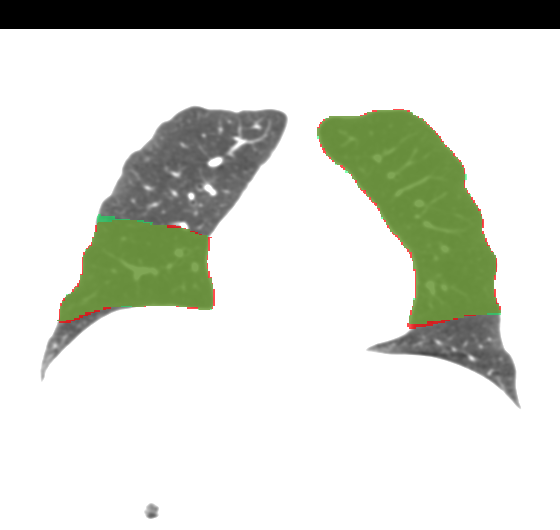}
  \end{subfigure}\hfill%
  \begin{subfigure}[b]{0.24\textwidth}
    \includegraphics[width=\textwidth]{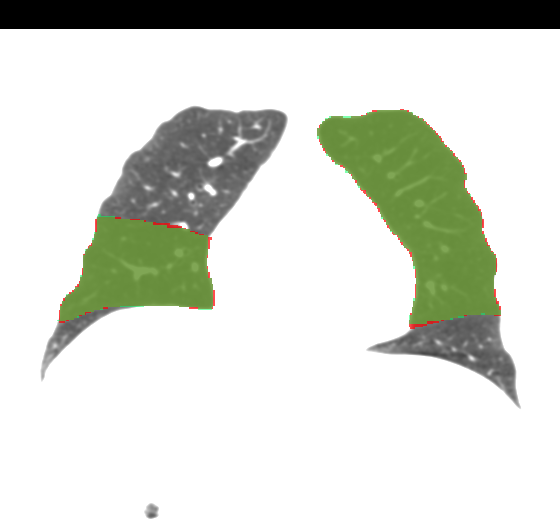}
  \end{subfigure}\hfill%

  \rotatebox{90}{\hspace{0.4cm}Lung lobes: Case 2}
  \begin{subfigure}[b]{0.24\textwidth}
    \includegraphics[width=\textwidth]{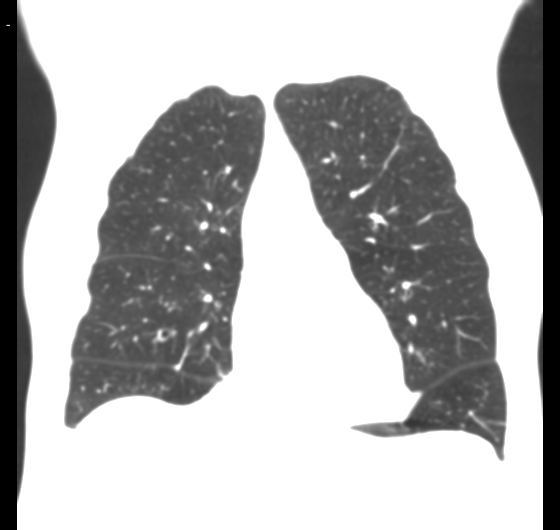}
    \caption{}
  \end{subfigure}\hfill%
  \begin{subfigure}[b]{0.24\textwidth}
    \includegraphics[width=\textwidth]{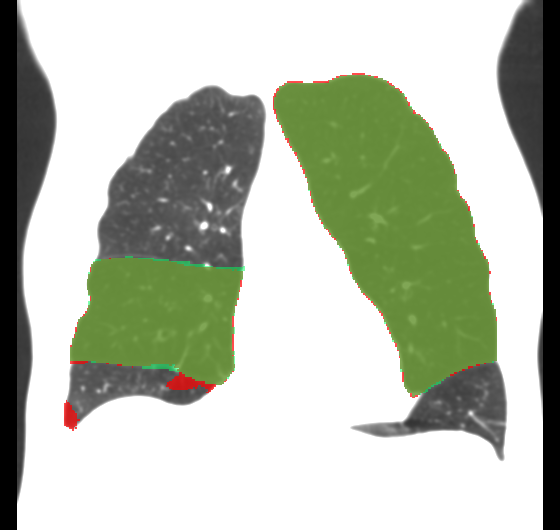}
    \caption{}
  \end{subfigure}\hfill%
  \begin{subfigure}[b]{0.24\textwidth}
    \includegraphics[width=\textwidth]{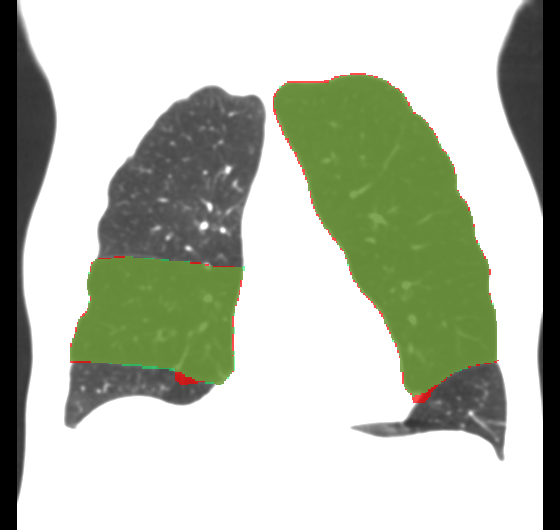}
    \caption{}
  \end{subfigure}\hfill%
  \begin{subfigure}[b]{0.24\textwidth}
    \includegraphics[width=\textwidth]{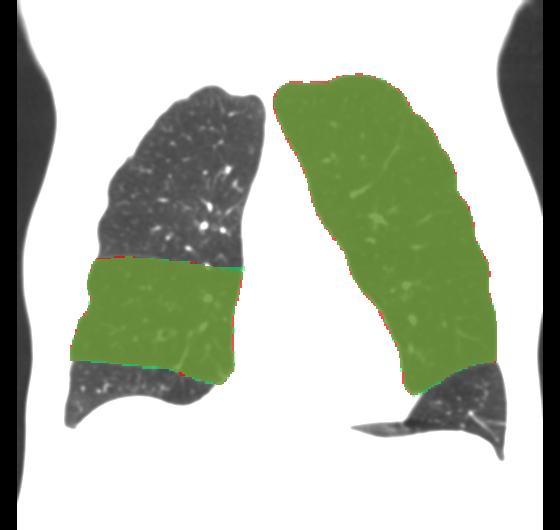}
    \caption{}
  \end{subfigure}\hfill%
  \caption{Predictions of the experiments of the lung lobes segmentation task trained on 30 CT scans. (a) Shows the original slice, and the training strategies (b) scratch \expe{LS30}, (c) vanilla transfer learning \expe{LT30}, and (d) transfer learning with fine-tuning \expe{LF30}.}
  \label{fig:predictionsset30lunglobes}
\end{figure*}
\section{Discussion}

This study evaluated transfer learning for segmentation tasks in medical imaging.
For this study, we focused on CT imaging and collected a large dataset consisting of 556 CT scans from 6 publicly available datasets. 
We trained a base segmentation model using this large but sparsely annotated dataset and evaluated whether transfer learning from the base model benefits four new segmentation tasks (esophagus, lung lobes, prostate, and vertebrae).
To utilize this limited annotated dataset, we trained the base model to learn from only the annotated regions and ignored other regions which could belong to one of the other classes (e.g., kidneys in scans in which only annotations of the liver are present). 
We analyzed different training strategies (training from scratch, vanilla transfer learning, and transfer learning with fine-tuning) and the influence of the training set size. We found that initializing a 3D U-Net with the learned parameters of the base model is beneficial, especially when only a limited number of annotated scans for the new task are available.
Models trained with small datasets (10 scans) that use transfer learning performed comparable with models trained from scratch with 40 or more scans. 
This finding aligns with previous research that showed that transfer learning reduces the need for large amounts of annotated data and obtains better performance than networks trained from scratch.
Most importantly, for datasets with up to 50 annotated images, transfer learning from the generic base model never hurt performance and can therefore be generally recommended. 
However, transfer learning might not be necessary if a larger dataset is available.
Nevertheless, it remains advantageous for reducing computational costs and mitigating carbon emissions due to faster convergence.
Note that the initial base model used in our experiments benefits from the multi-center data used to compose the large sparsely annotated dataset.
This dataset provides a diverse range of examples and may help the model generalize better to new challenging tasks.
For instance, the results of the lung lobes segmentation task (see \Cref{fig:predictionsset30lunglobes}) show how difficult it is to separate lung lobes.
While the prostate segmentation task (see \Cref{fig:predictionsset30prostate}) shows both transfer learning strategies obtained more consistent results than the experiment trained from scratch.
Note the predictions of the transfer learning with fine-tuning training strategy obtains better results than the vanilla transfer learning, except on the vertebrae segmentation task where the difference among training strategies is small.


Moreover, we investigated whether transfer learning from a model trained as proposed is more beneficial than transfer learning from a simpler model, trained for a single task and with a smaller dataset. 
While both transfer learning training strategies generally resulted in improved performance, especially for small training sets, transfer learning from a more generic model trained with many segmentation tasks consistently improved the performance, this was more task-dependent when using a single-task base model (\Cref{tab:vanillatlvertebrae2others}). 
The results presented in Tables \ref{tab:experiments} and \ref{tab:vanillatlvertebrae2others} provide insight into the performance of different transfer learning training strategies for organ segmentation tasks. 
Our findings suggest that vanilla transfer learning from the base model (Exp0) performs better than vanilla transfer learning from the simpler base model (\expe{VS90}), except for two experiments in the prostate segmentation task where the simpler base model slightly outperforms Exp0.
Furthermore, the simpler base model consistently performs better than experiments trained from scratch in all tasks except for the lung lobes segmentation task. 
This comparison suggests that transfer learning can be a useful approach for improving the performance of deep learning models in medical imaging, especially when annotated data is limited.
Although we trained the base model on CT scans, transfer learning from the base model (Exp0) was also favorable for the prostate segmentation task on MR images.
The simpler base model (\expe{VS90}) slightly obtained higher results in two experiments than the proposed base model.

While multi-center data is beneficial for generalization, the difference among annotation tools and protocols in datasets may difficult the learning process of the network.
For instance, \Cref{fig:predictionsset30esophagus} and \Cref{fig:predictionsset30prostate} show that the annotations (green regions) of the esophagus and prostate are not consistent in the sagittal orientation.
The annotation tool was designed to annotate the structures in a single orthogonal orientation, and no corrections were made on the other orthogonal orientations.
Moreover, all the networks learned to segment a continuous region.
Similarly, the dataset \gls{verse19} annotation protocol skips a partially present vertebra in the CT scan (see \Cref{fig:predictionsset30vertebrae}).

This study has a few limitations. 
First, some CT scans of the esophagus dataset were present in the large sparsely annotated dataset to train Exp0 but without the annotations of the esophagus. This data overlap may bias the transfer learning results for the esophagus segmentation task; other segmentation tasks (lung lobes, prostate, and vertebrae) contain unseen CT scans.
Moreover, this paper aims to provide a base model that benefits new medical segmentation tasks and not to obtain high performance on the large sparsely annotated dataset.

Second, although the difference in performance between experiments trained from scratch and using transfer learning was small when training with more than 50 CT scans, transfer learning experiments may be more robust in unseen images due to the knowledge learned from the multi-center dataset; future research should investigate whether this improved generalization is indeed observed.

Compared to our approach, \citet{Chen19a} created a binary prediction per each of the eight structures, while our approach uses a single softmax layer to obtain the predictions.
Federated learning has gained significant attention in recent years as an approach to facilitate multi-data center learning without the necessity of sharing sensitive patient data\citep{Xu23,Dong23,Shen22}.
It is noteworthy that while both federated learning and our study address sparsely annotated data, our study is not directly compatible with federated learning due to the centralization of data into a single large but sparsely annotated dataset.
This centralization approach, while effective for our research goal, differs from the decentralized nature of federated learning.

Overall, our study highlights the potential of transfer learning for organ segmentation in medical imaging, and our results provide valuable insights for researchers and practitioners looking to optimize the performance of deep learning models in this domain.
Further investigation may determine the optimal approach for selecting a base model and understanding the factors contributing to the performance differences observed in different tasks.
\section{Conclusions}
In conclusion, this study demonstrates the effectiveness of transfer learning in improving the performance of deep learning models in medical imaging, mainly when annotated data is limited.
Our results indicate that the learned features of a network trained on a partially annotated dataset can be transferred to new segmentation tasks, providing significant benefits, particularly on tasks where annotated data is scarce.
Our experiments show that transfer learning can be applied successfully to four segmentation tasks (esophagus, lung lobes, vertebrae, and prostate) and can significantly reduce the need for extensive annotation efforts.
Additionally, we have demonstrated that cross-modality transfer learning can be effective, as shown by our results in prostate segmentation in MR scans.
Furthermore, we found that fine-tuning the pre-trained base model before transfer learning is more beneficial than using vanilla transfer learning.
However, further research is needed to explore the limitations and potential applications of transfer learning in medical imaging and to develop more effective methods for utilizing sparsely annotated datasets.

\section*{Appendix}

\subsection*{Comparison of training strategies}
When comparing results per training strategy, the experiments that used transfer learning with fine-tuning obtained higher results than the other training strategies (scratch and vanilla transfer learning), except for the vertebrae experiments.
This improvement gradually reduces when increasing the training set size, see \Cref{fig:resultsmeandice}.
Transfer learning with fine-tuning was more beneficial to the esophagus experiments with small training sets, where the difference in Dice score between training strategies reached +0.129 Dice (\expe{EF10} 0.588 - \expe{ES10} 0.459) when training on ten scans.
Similarly, the prostate segmentation task experiments show that transfer learning is beneficial in most cases, except for \expe{PF10}, which performs similarly to \expe{PS10}.
The largest difference in the prostate segmentation reached +0.043 Dice (\expe{PF20} 0.863 - \expe{PS20} 0.819) when training on 20 scans.
While the lung lobes and vertebrae segmentation tasks show a slight improvement when using transfer learning with fine-tuning.

We compared two scenarios, networks trained on limited data and networks trained on large training data.
Our results show that transfer learning benefits experiments with limited data (usually up to 30 images).
Ideally, transfer learning would reduce the need for large training sets.
For instance, the difference between the experiment trained with 10 CT scans and the experiment trained from scratch on the full dataset of the esophagus decreased from +0.235 (\expe{ES158} - \expe{ES10}) to +0.106 Dice (\expe{ES158} - \expe{EF10}) after using transfer learning.
Moreover, the transfer learning experiment with fine-tuning reached 0.666 Dice when training with 40 images, reducing the difference to +0.027 in Dice score (\expe{ES158} - \expe{EF40}).
Note the transfer learning experiment reaches 0.027 difference in Dice score with 118 fewer images in the training set than \expe{ES158}.
For the lung lobes segmentation task, the performance difference between training with the entire training set and training from scratch using ten images reached +0.052 (\expe{LS90} 0.969 - \expe{LS10} 0.917).
Transfer learning with fine-tuning reduced that difference to +0.028 (\expe{LS90} 0.969 - \expe{LF10} 0.941).
Moreover, the transfer learning experiment with fine-tuning reached 0.961 Dice when training with 40 images, reducing the difference to 0.008 Dice (\expe{LS90} - \expe{LF40}).
Note the transfer learning experiment reaches a 0.008 difference in Dice score with 50 fewer images in the training set than \expe{LS90}.
Similarly, the difference for the vertebrae segmentation task reached +0.036 (\expe{VS90} 0.956 - \expe{VS10} 0.920).
The difference is already small; however, transfer learning with fine-tuning slightly reduced that difference to 0.027 (\expe{VS90} 0.956 - \expe{VF10} 0.929).
For this segmentation task, both transfer learning strategies obtained similar results.
The vanilla transfer learning experiment trained on 30 images reached a 0.942 Dice score, 0.014 less than the experiment trained from scratch on the full dataset; these experiments have a difference of 60 images in training sets.
Transfer learning did not benefit the prostate segmentation task with the smallest training set, as with previous segmentation tasks, but with more images in the training set.
For instance, when using 20 images as training set, the difference with the experiment trained from scratch on the full dataset decreased from +0.033 (\expe{PS45} 0.852 Dice - \expe{PS20} 0.819 Dice) to -0.01 (\expe{PS45} 0.852 Dice - \expe{PF20} 0.862 Dice) after using transfer learning. This shows that transfer learning with fine-tuning on 20 scans obtained already a higher performance than the experiment trained from scratch on the full dataset.
The difference in performance among training strategies when using the full dataset reached +0.032 (\expe{PF45} - \expe{PS45}) for the prostate segmentation task; this shows the benefits of cross-modality transfer learning. Note that the same difference in the CT segmentation tasks is lower (+0.0001 lung lobes, +0.015 esophagus, and -0.0012 vertebrae).
These results show that transfer learning was highly beneficial when having limited data, boosting performance and reducing the need for a large annotated dataset to obtain high segmentation performance.

Transfer learning increased the performance of segmentation tasks compared to experiments trained from scratch (see \Cref{tab:experiments} and \Cref{fig:resultsmeandice}).
Moreover, the experiments that use transfer learning with fine-tuning obtained higher results than vanilla transfer learning.
These results show that gradual training (fine-tuning) boosts the network's performance of the four new segmentation tasks, including the prostate segmentation (cross-modality transfer learning), which has MR data only.
\citet{Ragh19a} showed that different domain transfer learning does not improve performance.
This paper shows that the same domain/modality transfer learning is beneficial for the medical domain, including different modalities (CT to MR prostate).

\subsection*{Results of experiments trained on 10 scans}
\begin{figure*}[ht!]
  \rotatebox{90}{Esophagus: Case 1}
  \begin{subfigure}[b]{0.24\textwidth}
    \includegraphics[width=\textwidth]{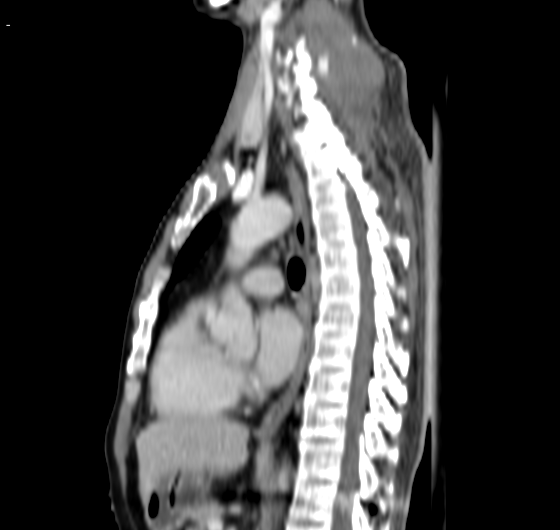}
  \end{subfigure}\hfill%
  \begin{subfigure}[b]{0.24\textwidth}
    \includegraphics[width=\textwidth]{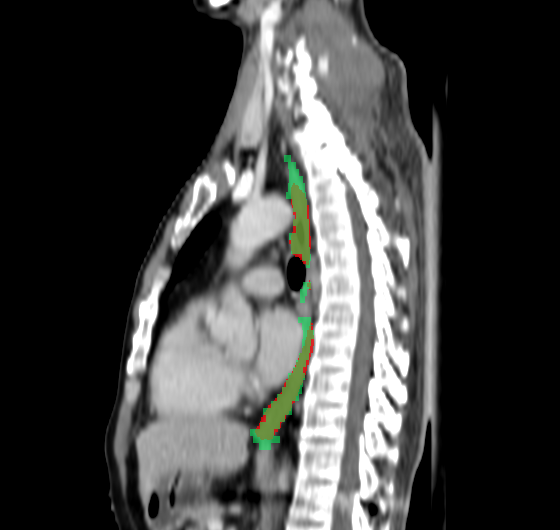}
  \end{subfigure}\hfill%
  \begin{subfigure}[b]{0.24\textwidth}
    \includegraphics[width=\textwidth]{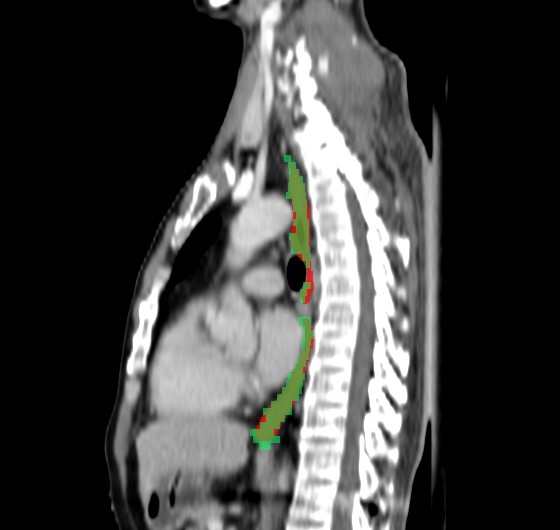}
  \end{subfigure}\hfill%
  \begin{subfigure}[b]{0.24\textwidth}
    \includegraphics[width=\textwidth]{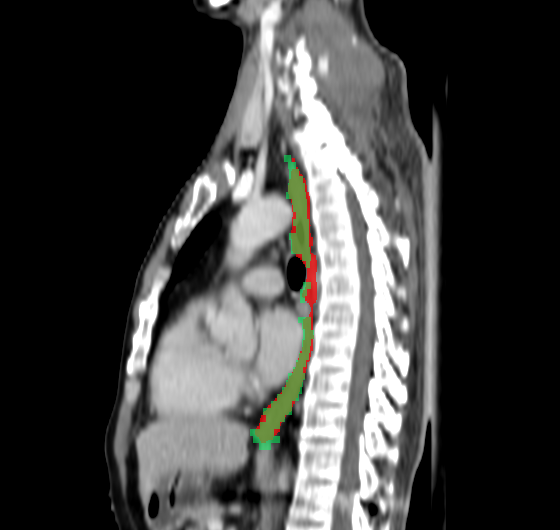}
  \end{subfigure}\hfill\vspace{0.25cm}
  
  \rotatebox{90}{Esophagus: Case 2}
  \begin{subfigure}[b]{0.24\textwidth}
    \includegraphics[width=\textwidth]{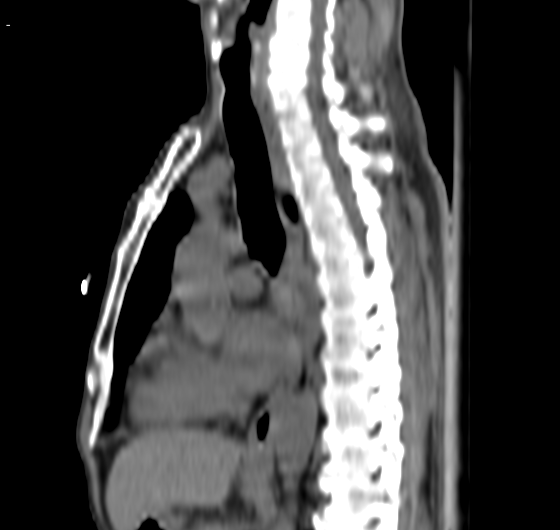}
  \end{subfigure}\hfill%
  \begin{subfigure}[b]{0.24\textwidth}
    \includegraphics[width=\textwidth]{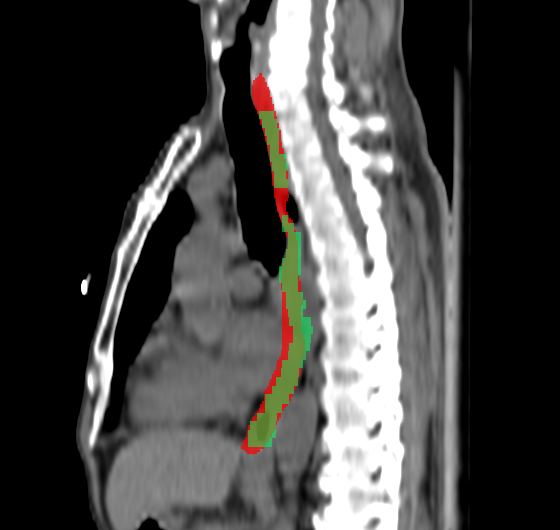}
  \end{subfigure}\hfill%
  \begin{subfigure}[b]{0.24\textwidth}
    \includegraphics[width=\textwidth]{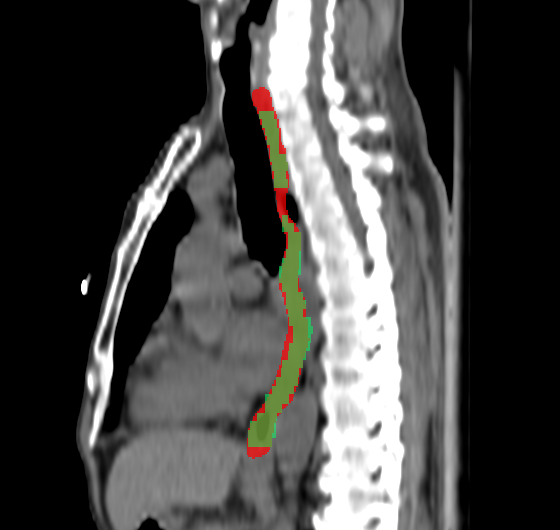}
  \end{subfigure}\hfill%
  \begin{subfigure}[b]{0.24\textwidth}
    \includegraphics[width=\textwidth]{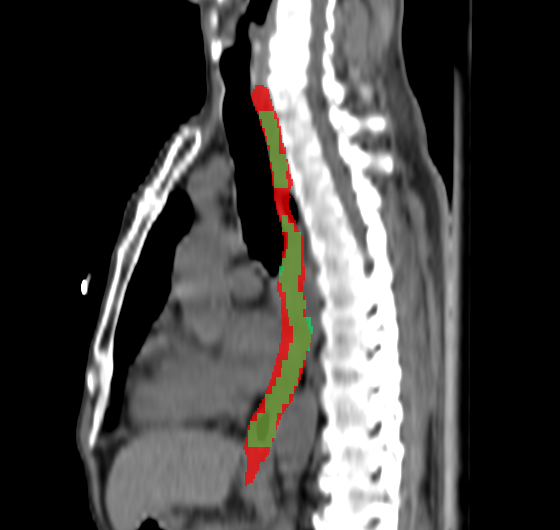}
  \end{subfigure}\hfill\vspace{0.25cm}

  \rotatebox{90}{\hspace{0.5cm}Esophagus: Case 3}
  \begin{subfigure}[b]{0.24\textwidth}
    \includegraphics[width=\textwidth]{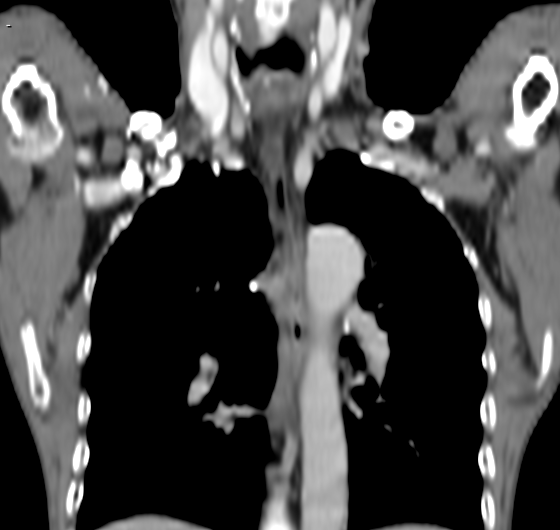}
    \caption{}
  \end{subfigure}\hfill%
  \begin{subfigure}[b]{0.24\textwidth}
    \includegraphics[width=\textwidth]{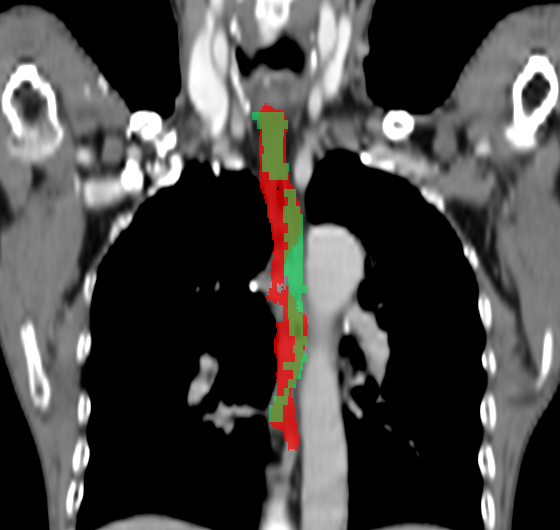}
    \caption{}
  \end{subfigure}\hfill%
  \begin{subfigure}[b]{0.24\textwidth}
    \includegraphics[width=\textwidth]{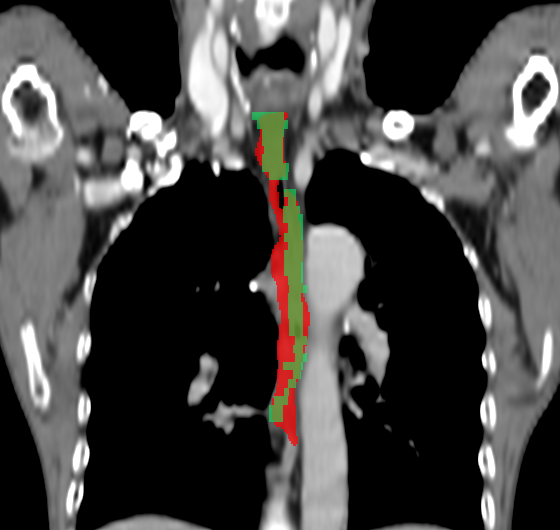}
    \caption{}
  \end{subfigure}\hfill%
  \begin{subfigure}[b]{0.24\textwidth}
    \includegraphics[width=\textwidth]{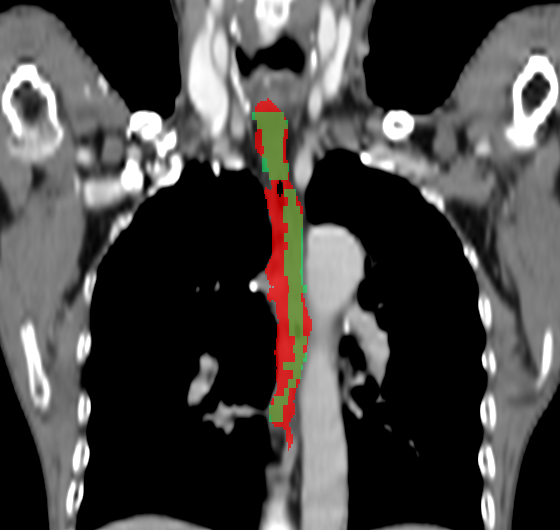}
    \caption{}
  \end{subfigure}\hfill%
  \caption{Predictions of the experiments of the esophagus segmentation task trained on 10 CT scans. (a) Shows the original slice, and the training strategies (b) scratch \expe{ES10}, (c) vanilla transfer learning \expe{ET10}, and (d) transfer learning with fine-tuning \expe{EF10}.}
  \label{fig:predictionsset10esophagus}
\end{figure*}

\begin{figure*}[ht!]
  \rotatebox{90}{Vertebrae: Case 1}
  \begin{subfigure}[b]{0.24\textwidth}
    \includegraphics[width=\textwidth]{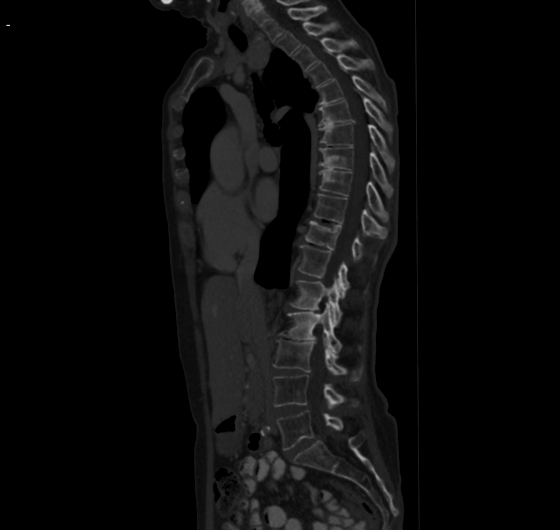}
  \end{subfigure}\hfill%
  \begin{subfigure}[b]{0.24\textwidth}
    \includegraphics[width=\textwidth]{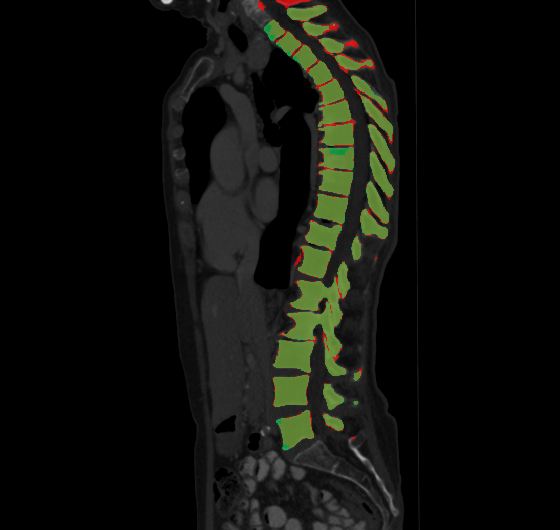}
  \end{subfigure}\hfill%
  \begin{subfigure}[b]{0.24\textwidth}
    \includegraphics[width=\textwidth]{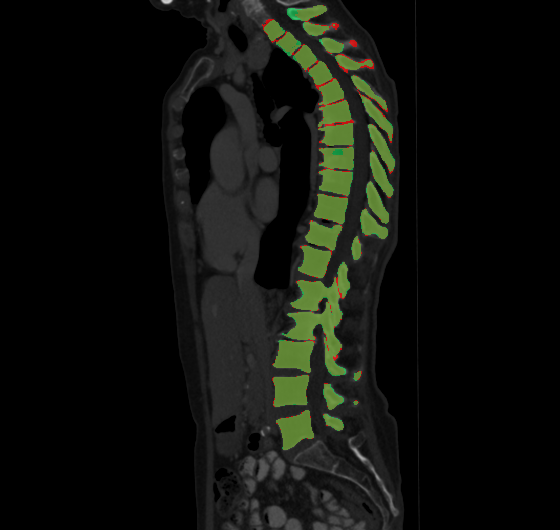}
  \end{subfigure}\hfill%
  \begin{subfigure}[b]{0.24\textwidth}
    \includegraphics[width=\textwidth]{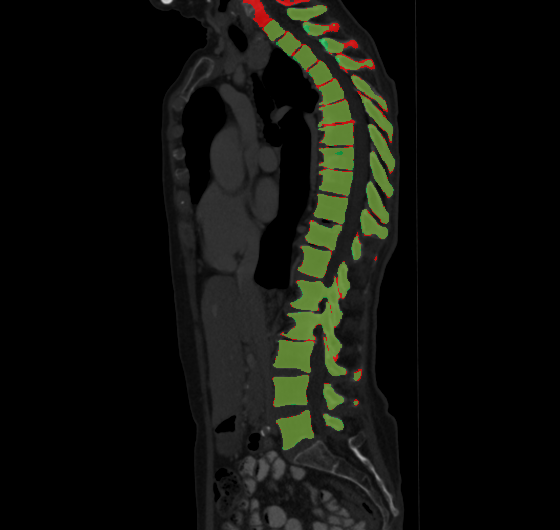}
  \end{subfigure}\hfill\vspace{0.25cm}

  \rotatebox{90}{\hspace{0.5cm}Vertebrae: Case 2}
  \begin{subfigure}[b]{0.24\textwidth}
    \includegraphics[width=\textwidth]{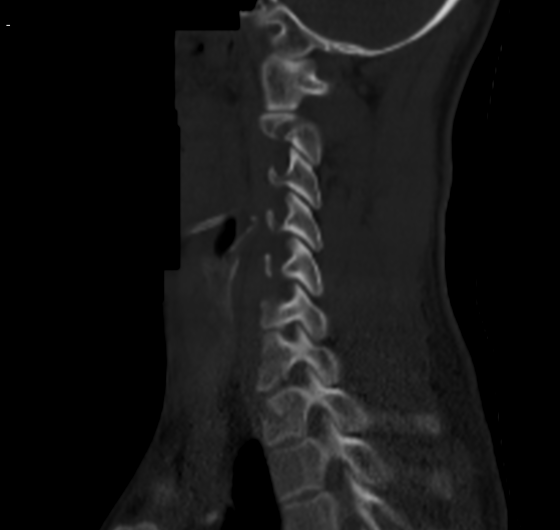}
    \caption{}
  \end{subfigure}\hfill%
  \begin{subfigure}[b]{0.24\textwidth}
    \includegraphics[width=\textwidth]{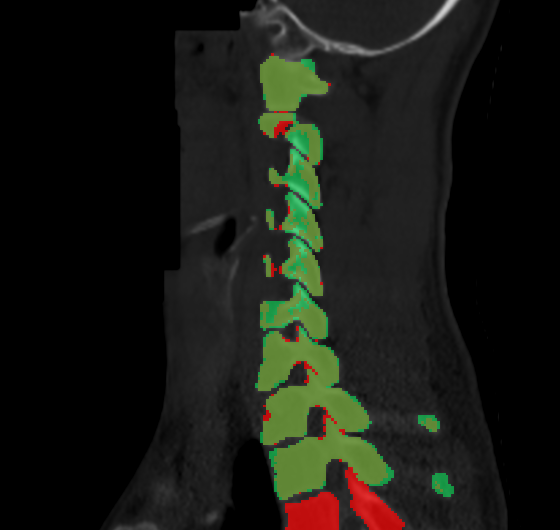}
    \caption{}
  \end{subfigure}\hfill%
  \begin{subfigure}[b]{0.24\textwidth}
    \includegraphics[width=\textwidth]{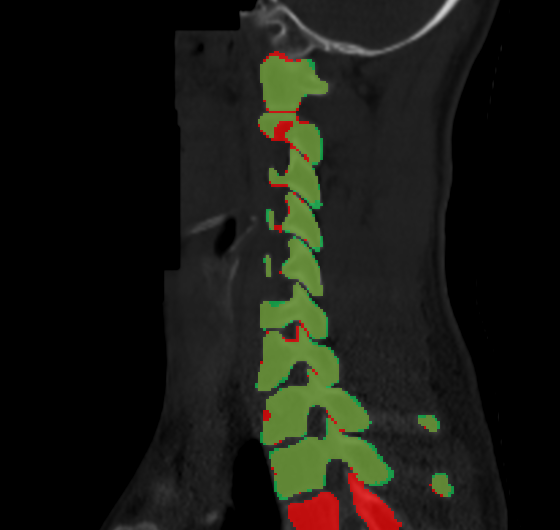}
    \caption{}
  \end{subfigure}\hfill%
  \begin{subfigure}[b]{0.24\textwidth}
    \includegraphics[width=\textwidth]{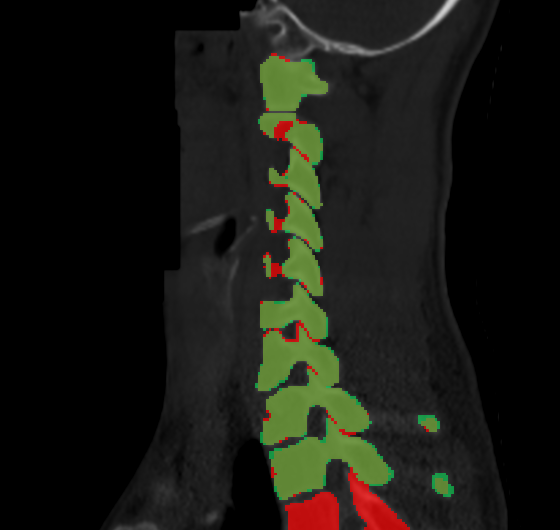}
    \caption{}
  \end{subfigure}\hfill%
  \caption{Predictions of the experiments of the vertebrae segmentation task trained on 10 CT scans. (a) Shows the original slice, and the training strategies (b) scratch \expe{VS10}, (c) vanilla transfer learning \expe{VT10}, and (d) transfer learning with fine-tuning \expe{VF10}.}
  \label{fig:predictionsset10vertebrae}
\end{figure*}

\begin{figure*}[ht!]
  \rotatebox{90}{Prostate: Case 1}
  \begin{subfigure}[b]{0.24\textwidth}
    \includegraphics[width=\textwidth]{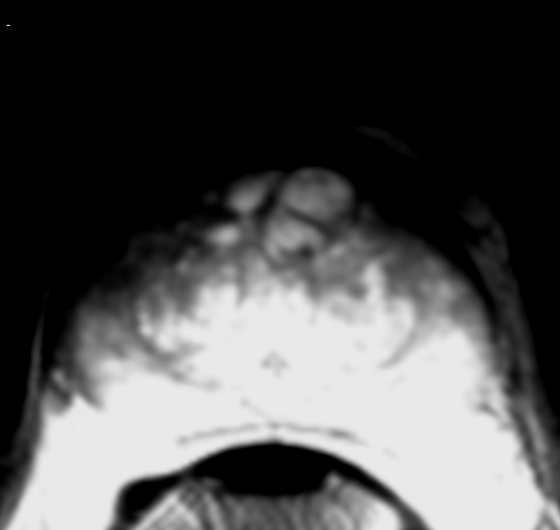}
  \end{subfigure}\hfill%
  \begin{subfigure}[b]{0.24\textwidth}
    \includegraphics[width=\textwidth]{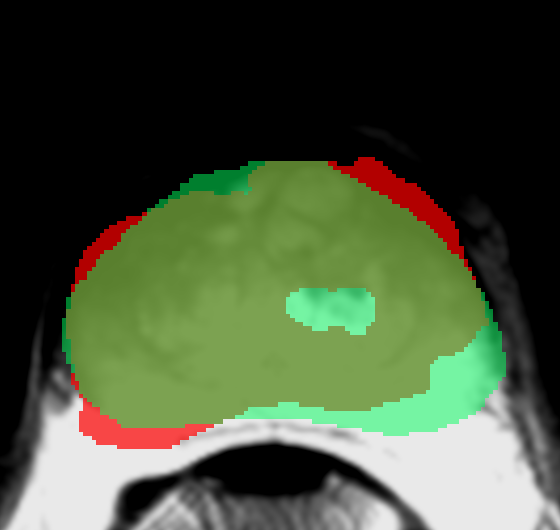}
  \end{subfigure}\hfill%
  \begin{subfigure}[b]{0.24\textwidth}
    \includegraphics[width=\textwidth]{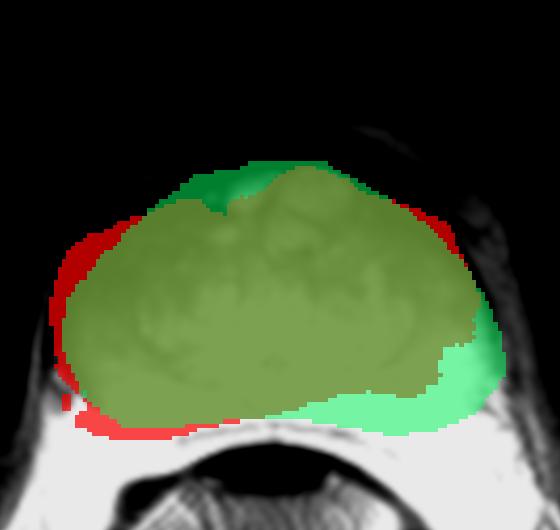}
  \end{subfigure}\hfill%
  \begin{subfigure}[b]{0.24\textwidth}
    \includegraphics[width=\textwidth]{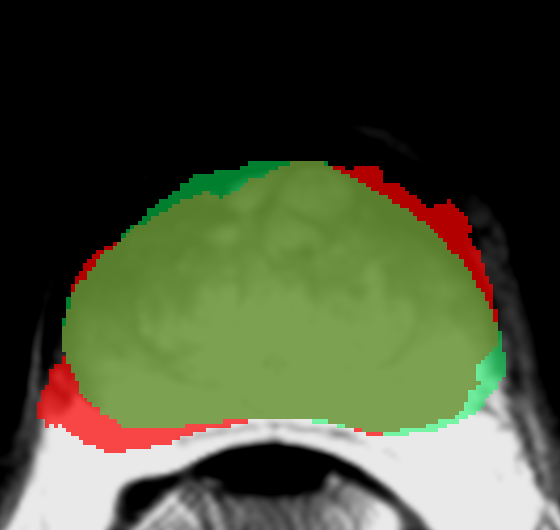}
  \end{subfigure}\hfill\vspace{0.25cm}

  \rotatebox{90}{\hspace{0.5cm}Prostate: Case 2}
  \begin{subfigure}[b]{0.24\textwidth}
    \includegraphics[width=\textwidth]{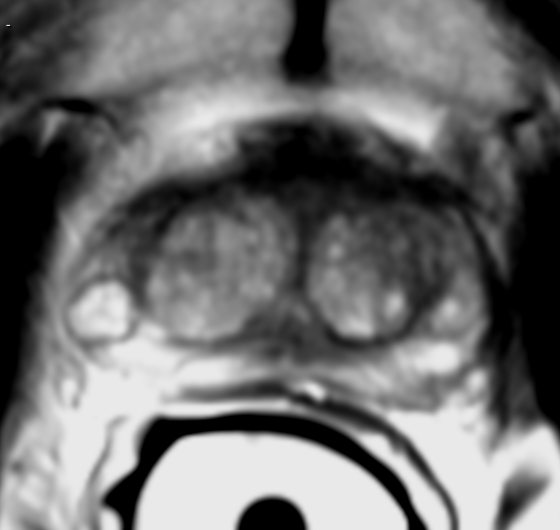}
    \caption{}
  \end{subfigure}\hfill%
  \begin{subfigure}[b]{0.24\textwidth}
    \includegraphics[width=\textwidth]{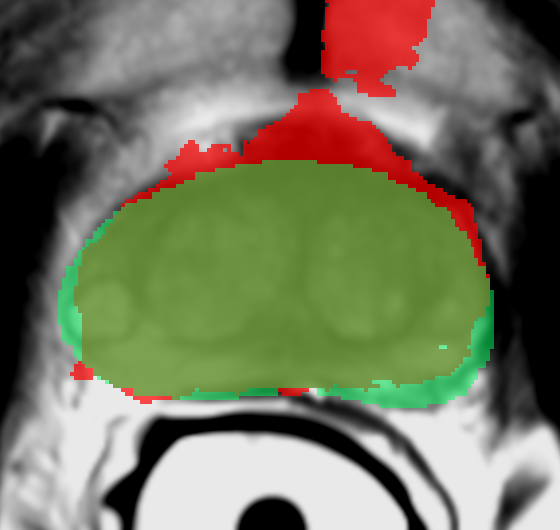}
    \caption{}
  \end{subfigure}\hfill%
  \begin{subfigure}[b]{0.24\textwidth}
    \includegraphics[width=\textwidth]{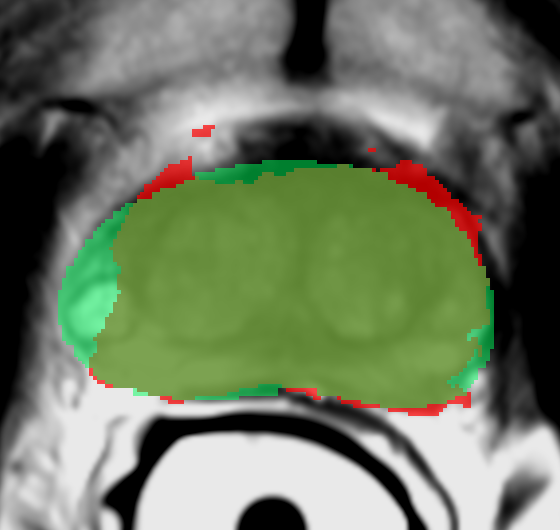}
    \caption{}
  \end{subfigure}\hfill%
  \begin{subfigure}[b]{0.24\textwidth}
    \includegraphics[width=\textwidth]{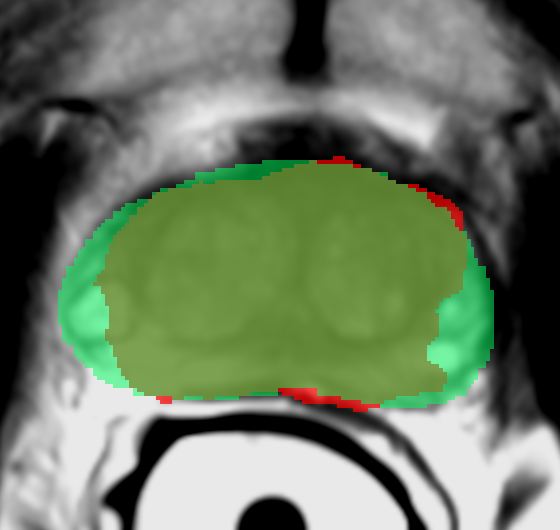}
    \caption{}
  \end{subfigure}\hfill%
  \caption{Predictions of the experiments of the prostate segmentation task trained on 10 CT scans. (a) Shows the original slice, and the training strategies (b) scratch \expe{PS10}, (c) vanilla transfer learning \expe{PT10}, and (d) transfer learning with fine-tuning \expe{PF10}.}
  \label{fig:predictionsset10prostate}
\end{figure*}

\begin{figure*}[ht!]
  \rotatebox{90}{Lung lobes: Case 1}
  \begin{subfigure}[b]{0.24\textwidth}
    \includegraphics[width=\textwidth]{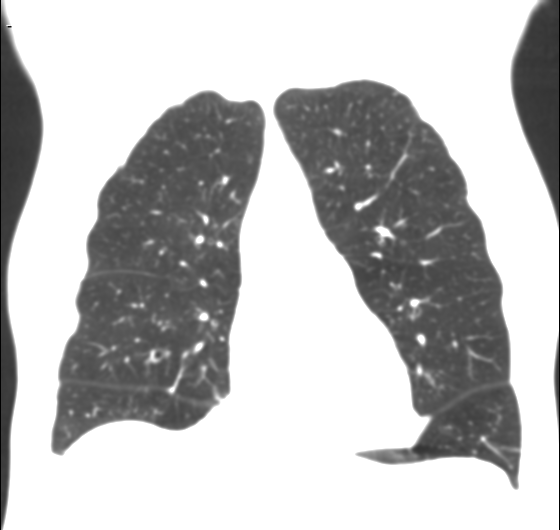}
  \end{subfigure}\hfill%
  \begin{subfigure}[b]{0.24\textwidth}
    \includegraphics[width=\textwidth]{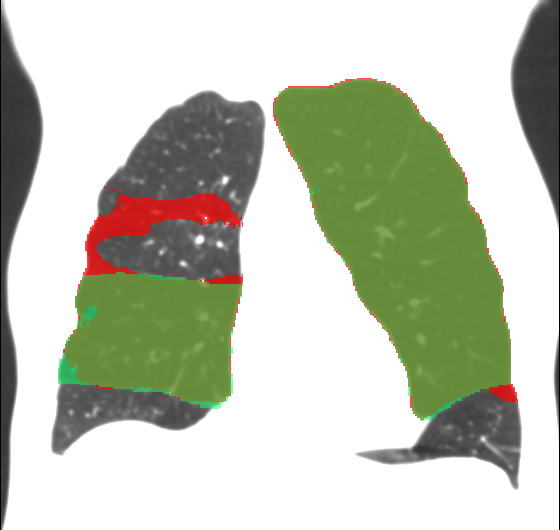}
  \end{subfigure}\hfill%
  \begin{subfigure}[b]{0.24\textwidth}
    \includegraphics[width=\textwidth]{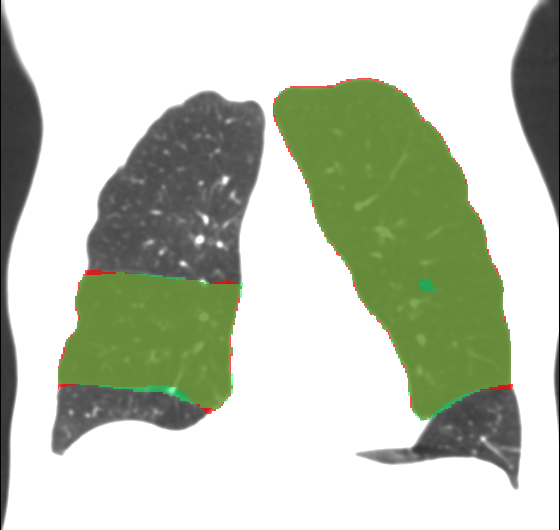}
  \end{subfigure}\hfill%
  \begin{subfigure}[b]{0.24\textwidth}
    \includegraphics[width=\textwidth]{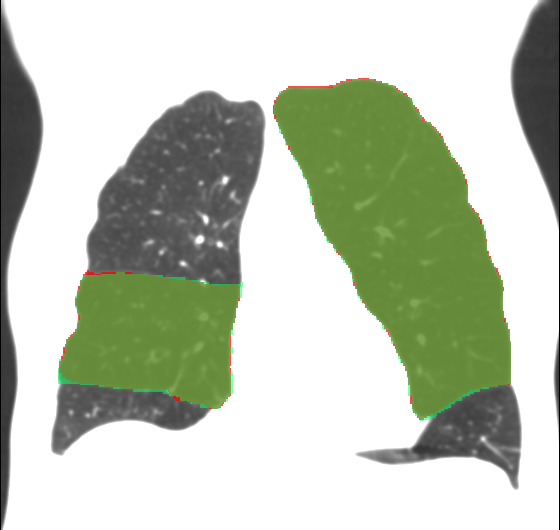}
  \end{subfigure}\hfill\vspace{0.25cm}

  \rotatebox{90}{Lung lobes: Case 2}
  \begin{subfigure}[b]{0.24\textwidth}
    \includegraphics[width=\textwidth]{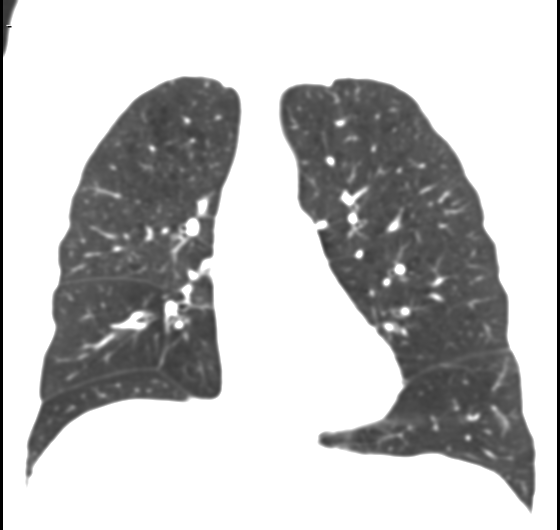}
  \end{subfigure}\hfill%
  \begin{subfigure}[b]{0.24\textwidth}
    \includegraphics[width=\textwidth]{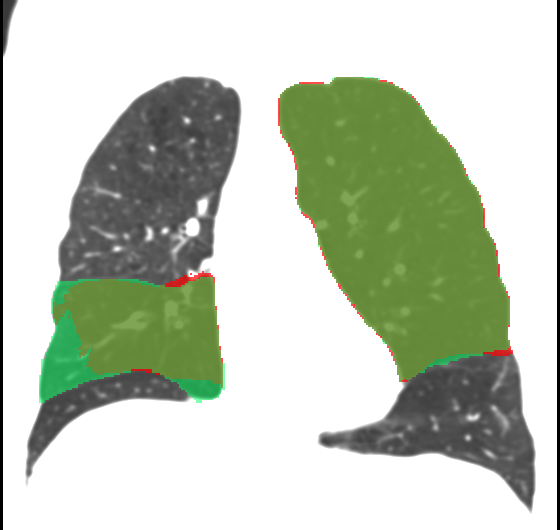}
  \end{subfigure}\hfill%
  \begin{subfigure}[b]{0.24\textwidth}
    \includegraphics[width=\textwidth]{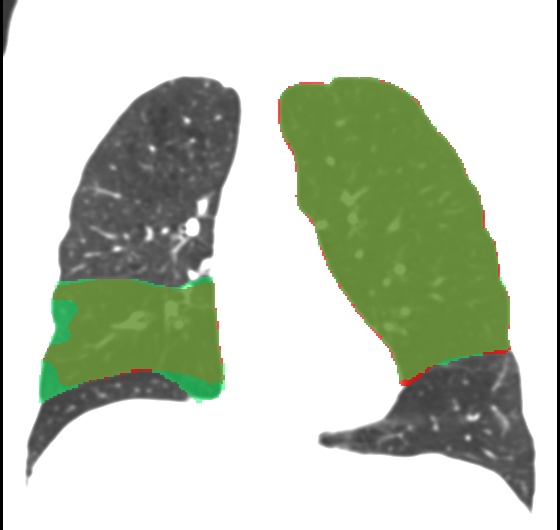}
  \end{subfigure}\hfill%
  \begin{subfigure}[b]{0.24\textwidth}
    \includegraphics[width=\textwidth]{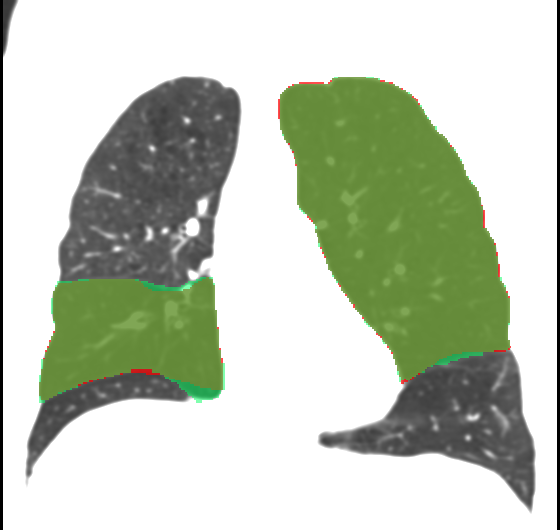}
  \end{subfigure}\hfill\vspace{0.25cm}

  \rotatebox{90}{\hspace{0.3cm}Lung lobes: Case 3}
  \begin{subfigure}[b]{0.24\textwidth}
    \includegraphics[width=\textwidth]{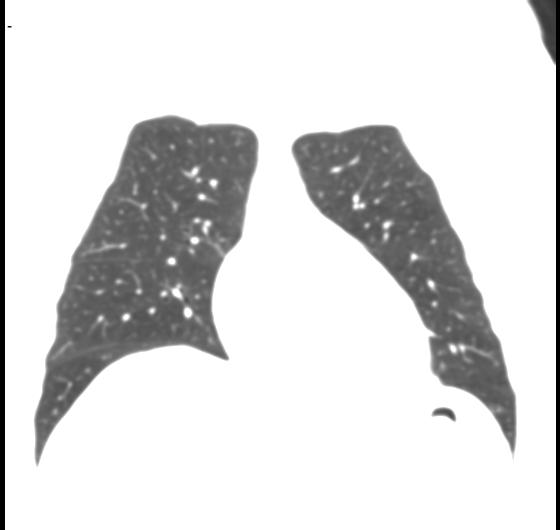}
    \caption{}
  \end{subfigure}\hfill%
  \begin{subfigure}[b]{0.24\textwidth}
    \includegraphics[width=\textwidth]{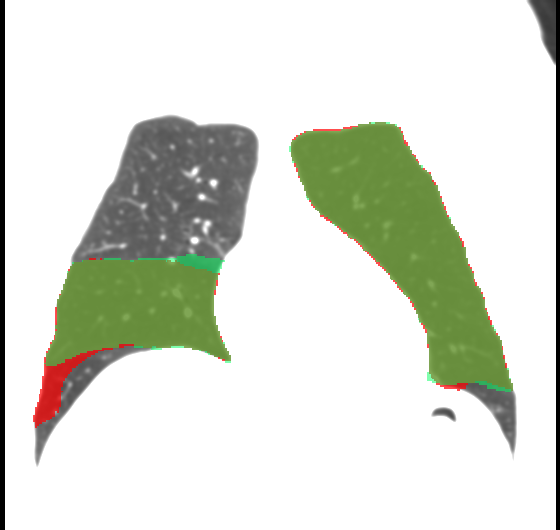}
    \caption{}
  \end{subfigure}\hfill%
  \begin{subfigure}[b]{0.24\textwidth}
    \includegraphics[width=\textwidth]{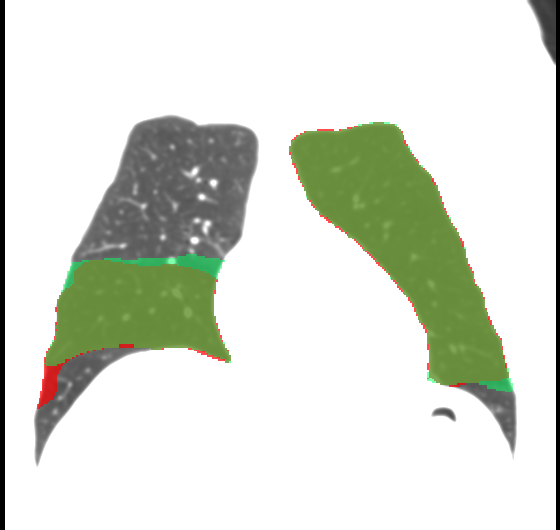}
    \caption{}
  \end{subfigure}\hfill%
  \begin{subfigure}[b]{0.24\textwidth}
    \includegraphics[width=\textwidth]{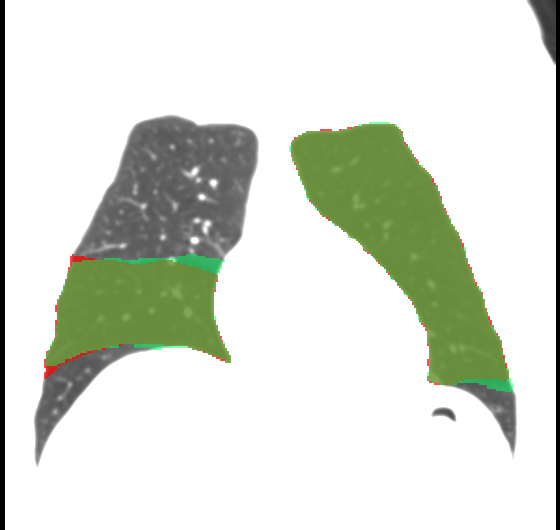}
    \caption{}
  \end{subfigure}\hfill%
  \caption{Predictions of the experiments of the lung lobes segmentation task trained on 10 CT scans. (a) Shows the original slice, and the training strategies (b) scratch \expe{LS10}, (c) vanilla transfer learning \expe{LT10}, and (d) transfer learning with fine-tuning \expe{LF10}.}
  \label{fig:predictionsset10lunglobes}
\end{figure*}

This section shows the results of the experiments trained on 10 scans. This figures are \ref{fig:predictionsset10esophagus}, \ref{fig:predictionsset10prostate}, \ref{fig:predictionsset10lunglobes}, and \ref{fig:predictionsset10vertebrae}.

\bibliographystyle{unsrtnat}

\end{document}